\documentclass[prb,preprint]{revtex4}
\usepackage{graphicx}
\makeatletter
\usepackage{epsfig}
\def\@dotsep{4.5}
\usepackage{dcolumn}
\usepackage{amsmath}
\makeatother

\begin{document}

\title{Diffraction of fast atoms and molecules from surfaces}

\author{ J. R. Manson}
 \affiliation{ Department of Physics and Astronomy, Clemson University, Clemson,
SC, 29634, USA}
 \email{jmanson@clemson.edu}   

\author{ Hocine Khemliche and Philippe Roncin}
 \affiliation{ Laboratoire des Collisions Atomiques et Mol\'eculaires \\
Universit\'e de Paris-Sud \\ 91405 Orsay, France}

 \email{jmanson@clemson.edu}

\date{\today}

\begin{abstract}
Prompted by recent experimental developments, a theory of surface scattering of fast atoms at grazing incidence is developed.
The theory gives rise to a quantum mechanical limit for ordered surfaces that describes coherent diffraction peaks whose thermal attenuation is governed by a Debye-Waller factor, however, this Debye-Waller factor has values much larger than would be calculated using simple models.
A classical limit for incoherent scattering is obtained for high energies and temperatures.  Between these limiting classical and quantum cases is another regime in which diffraction features appear that are broadened by the motion in the fast direction of the scattered beam but whose intensity is not governed by a Debye-Waller factor.
All of these limits appear to be accessible within the range of currently available experimental conditions.
\end{abstract}

\maketitle


\section{Introduction}

Recently, two laboratories have demonstrated that it is possible to diffract beams of fast-moving atoms and molecules from ordered surfaces under conditions of grazing incident angles, a method that has been given the name Grazing Incidence Fast Atom Diffraction (GIFAD).~\cite{Roncin-07,Winter-07,Winter-08} The range of energies goes from several hundred eV up to tens of keV  for the projectiles used which include He, H atoms and H$_2$ molecules.  The target surfaces were single-crystal alkali halides.  These energies correspond to de Broglie wavelengths several orders of magnitude smaller than the lattice spacing of the target crystal.  Because of the long travel path near the surface one might expect sufficient dephasing to occur which would destroy all quantum coherence, and furthermore a simple application of the Debye-Waller attenuation due to thermal roughness of the surface would imply totally negligible diffraction intensities except possibly for the most grazing incidence conditions.  However, when the incident beam was aligned along a high symmetry direction intense and clearly measurable diffraction peak signals were observed in the direction normal to the scattering plane.

Two different theoretical groups have carried out numerical simulations of the diffraction process using propagation of wave packet techniques and have shown that one can obtain a reasonable description of the scattering intensities as functions of the initial beam parameters such as energy and incident angle.~\cite{Roncin-07,Burgdorfer-07}  These descriptions indicate that along the fast velocity component of the incident beam the interaction potential of the surface is averaged so that the total potential appears as if it were relatively smooth with one-dimensional corrugations in the high symmetry direction.  Also, it has been shown that simple calculations of diffraction intensities based on the eikonal approximation for a one-dimensional repulsive hard wall potential can predict the diffraction intensities reasonably well.~\cite{Roncin-07}

In this paper a theory is developed starting from semiclassical quantum mechanics that produces analytical expressions which describe the scattering process at levels of coherence ranging from purely quantum mechanical diffraction to the fully incoherent classical limit.  The basic starting approximations are the eikonal approximation and assumption of a target crystal whose thermal vibrations are in the harmonic limit.  The scattering process is described as a series of sequential collisions along the path of the projectile as it moves near the crystal.  Under these conditions very little energy is lost due to motion in the fast direction parallel to the surface and it is shown that the Debye-Waller factor becomes sufficiently large to permit measurable diffraction intensities.
Under conditions where quantum peaks are observed the very small energy loss associated with the fast direction ensures that the dephasing is sufficiently weak to permit quantum scattering in the other two mutually perpendicular directions.  In addition to the purely quantum diffraction limit the theory can be taken to the classical limit through use of the Bohr correspondence principle. In between the clearly distinct quantum and classical limits there is a regime where the dephasing allows for distinct quantum peak features that are broadened by a limited coherence length and these quantum diffraction features do not obey Debye-Waller attenuation.  Each of these regimes has its own signature characteristics that should be measurable in the energy, angular and temperature dependence of the measurements.

In the remainder of this paper the theory is developed together
 with a description of the various possible quantum mechanical and classical limits in the following Section~\ref{theory}.
In Sec.~\ref{calcs} some numerically
calculated results are shown and discussed.
Conclusions are
discussed in Sec.~\ref{conclusion}.
Some of the results used in developing the various models of Sec.~\ref{theory} are developed in the Appendices.

\section{Theory}\label{theory}

An appropriate starting point for developing the theory of scattering of atomic projectiles from a surface is the quantum mechanical transition rate for a particle of incident wave vector ${\bf k}_i$ making a transition to the final state of wave vector ${\bf k}_f$ given by
\begin{eqnarray} \label{golden}
w({\bf k}_f,{\bf k}_i) ~=~ \frac{2 \pi}{\hbar}
\left\langle   \sum_{\{ n_f \}}   |T_{fi}|^2 \delta({\cal E}_f - {\cal E}_i)       \right \rangle
~,
\end {eqnarray}
where $ T_{fi}$ is the transition matrix element taken with respect to the final and initial states of the  system of projectile plus target, $ {\cal E}_i$ and  $ {\cal E}_f$ are the initial and final global energies of the entire system.  The angular brackets indicate an average over all initial states of the surface target and the $ \sum_{\{ n_f \}} $ indicates a sum over all final states of the target.

The usual procedure is to write the energy delta-function as its time-integral Fourier representation and to shift to the interaction picture in which the time dependence is governed by the target hamiltonian~\cite{Weinstock-44}, a process that is sometimes called the Glauber-Van Hove transformation~\cite{Glauber-52,Hove-54}:
\begin{eqnarray} \label{Glauber}
w({\bf k}_f,{\bf k}_i) ~=~ \frac{1}{\hbar^2} ~ \int_{-\infty}^{+\infty} \, d t ~
e^{i (E_i-E_f)t/\hbar} ~
\left\langle  T_{if}(t)  T_{fi}(0)     \right\rangle
~,
\end {eqnarray}
where $E_f$ and $E_i$ are the final and initial translational energies of the projectile atom.

The fast atom diffraction experiment is one in which one component of the incident momentum, chosen as $\hbar k_{ix}$,  is much larger than the component in the direction normal to the surface which is chosen to be the $z$-direction.  Thus, it will be a reasonable approximation to assume that the transition operator separates into the product of an operator in the normal direction and an operator in the directions parallel to the surface
\begin{eqnarray} \label{T}
\hat{T} ~=~ \hat{T}_{z} ~ \hat{T}_{xy}
~.
\end {eqnarray}
The matrix element of the transition operator, taken with respect to final and initial states of the projectile, then separates into the product
\begin{eqnarray} \label{T2}
T_{fi} ~=~ \left(  \Phi_f(z) \left| \hat{T}_{z} \right| \Phi_i(z) \right)  ~
\left(  \Psi_f(x,y) \left| \hat{T}_{xy} \right| \Psi_i(x,y) \right)
~.
\end {eqnarray}

The squared normal transition operator $ \left|\left(  \Phi_f(z) \left| \hat{T}_{z}  \right|\Phi_i(z)\right) \right|^2$ is proportional to the probability of making a transition from $k_{iz}$ to  $k_{fz}$.  There are two useful limits in which this can be evaluated exactly.  The first is the nearly trivial case of a surface whose potential is a rigid hard repulsive wall in which case it is a constant.  This is the evaluation used in the standard application of the eikonal approximation for elastic scattering.~\cite{Levi-77}
The other useful limit is given by the classical motion of a grazing-angle projectile moving in a one-dimensional exponentially repulsive potential such as
\begin{eqnarray} \label{ex}
V(z) ~=~ V_0 ~ e^{-\Gamma z}
~.
\end {eqnarray}
If the atoms of the surface are subject to small random displacements about their equilibrium positions the scattered distribution
 becomes a lognormal distribution in the final angle~\cite{Roncin,Villette} as shown below in Appendix~\ref{lognorm}:
\begin{eqnarray} \label{lognormal}
\left|\left(  \Phi_f(z) \left| \hat{T}_{z} \right| \Phi_i(z) \right) \right|^2 ~
\propto ~ P(\theta_f) ~=~ \sqrt{\frac{2}{\pi}} \frac{1}{\Gamma \sigma \theta_f} ~
\exp \left\{ -\frac{2}{\Gamma^2 \sigma^2} \left[ \ln\left( \frac{~~\theta_f~~}{\overline{\theta_f}}  \right)    \right]^2 \right\}
~,
\end {eqnarray}
where $\sigma^2$ is the mean square displacement.
The origin of the surface displacement is thermal motion, although the displacement appears static to the fast-moving projectiles because the collision times are much shorter than typical phonon periods.
Thus $\sigma^2$ is given by the mean square
thermal displacement normal to the surface, which in the high temperature limit for a Debye model of the phonon density of states is
\begin{eqnarray} \label{sig}
\sigma^2 ~=~ \left\langle  u_z^2   \right\rangle  ~=~
\frac{3 \hbar^2 T_S}{M_C k_B \Theta_D^2}
~,
\end {eqnarray}
where $k_B$ is Boltzmann's constant, $T_S$ is the surface temperature, $M_C$ is the surface mass and $\Theta_D$ is the surface Debye temperature.

The matrix elements of the operator $\hat{T}_{xy} $ will be evaluated semiclassically within the eikonal approximation following the work of Bortolani and Levi~\cite{Bortolani}.   For purely elastic scattering from a rigid potential the  wave function in the asymptotic region far from the surface is of the form
\begin{eqnarray} \label{asym}
\psi_i({\bf r}) ~=~ e^{i {\bf k}_i \cdot {\bf r}} ~-~
\sum_{{\bf K}} ~ A({\bf K}) ~ e^{i ({\bf K}_i + {\bf K}) \cdot {\bf R}} ~
e^{i k_{fz} z}
~,
\end {eqnarray}
where $ {\bf K}_i$ is the surface-parallel component of the incident wave vector ${\bf k}_i$ and ${\bf K}={\bf K}_f-{\bf K}_i $.  The perpendicular component $ k_{fz}$ of ${\bf k}_f$ is determined by energy conservation $ E_f=E_i$ which implies
\begin{eqnarray} \label{circ}
k_{fz}^2 = k_{iz}^2 - K^2 - 2 {\bf K} \cdot {\bf K}_i
~.
\end {eqnarray}
The coefficient $A({\bf K}) $ of the outgoing asymptotic wave is related to the transition matrix element through~\cite{MansonSpringer}
\begin{eqnarray} \label{A}
\left(  \Psi_f(x,y) \left| \hat{T}_{xy} \right| \Psi_i(x,y) \right) ~=~
i e^{i \delta_f} ~ \frac{\hbar^2 k_{fz}}{mL} ~
 A({\bf K})
~,
\end {eqnarray}
where $L$ is the quantization length.

The eikonal approximation consists in first applying the Rayleigh ansatz, i.e., assuming that the wave function of Eq.~(\ref{asym}) is valid near the surface and then applying the correct hard wall boundary condition
\begin{eqnarray} \label{asymBC}
\psi({\bf R}, z=\xi({\bf R})) ~=~ 0
~,
\end {eqnarray}
where $\xi({\bf R}) $ is the corrugation function of the repulsive wall.
The eikonal approximation is obtained by further assuming that $ k_{fz} $ is weakly dependent on ${\bf K}$ which allows for a simple evaluation of the coefficient of the outgoing wave
\begin{eqnarray} \label{A1}
 A({\bf K}) ~=~  \frac{1}{L^2}~ \int ~d{\bf R} ~ e^{-i {\bf K} \cdot {\bf R}} ~
e^{- i \Delta k_{z} \xi({\bf R})}
~,
\end {eqnarray}
where $\Delta k_{z} = k_{iz} + k_{fz}$.

Eq.~(\ref{A1}) is the eikonal solution to the purely elastic problem.  The more general problem of a potential with thermal vibrations can be treated by introducing the displacement function $ {\bf u}({\bf R},t) $ of the surface into the coefficient
$ A({\bf K})$ using the transformations $ {\bf R} \longrightarrow {\bf R} - {\bf u}({\bf R},t)$ and $\xi({\bf R}) \longrightarrow \xi({\bf R}) - { u}_z({\bf R},t)$.    Applying this transformation to the transition matrix through Eq.~(\ref{A1}) casts the transition rate of Eq.~(\ref{Glauber}) into the form
\begin{eqnarray} \label{Glauber2}
w({\bf k}_f,{\bf k}_i) ~=~ \frac{1}{\hbar^2} ~ \left(\frac{\hbar^2 k_{fz}}{mL}\right)^2
~ \left|\left(  \Phi_f(z) \left| \hat{T}_{z}  \right|\Phi_i(z)\right) \right|^2
~ \int_{-\infty}^{+\infty} \, d t ~
e^{i (E_i-E_f)t/\hbar} ~
\\ \nonumber \times ~~~
\frac{1}{L^4} ~ \int ~d {\bf R} ~ \int ~d {\bf R^\prime} ~
e^{-i {\bf K} \cdot ({\bf R}-{\bf R^\prime})} ~
e^{-i \delta k_z [\xi({\bf R}) -\xi({\bf R^\prime}) ]}
\left\langle  e^{i \Delta{\bf  k} \cdot {\bf u}({\bf R},t)}
 e^{-i \Delta{\bf  k} \cdot {\bf u}({\bf R^\prime},0)}
\right\rangle
~,
\end {eqnarray}
where $\Delta{\bf  k} = {\bf k}_f - {\bf k}_i = ({\bf K}, \Delta k_{z})$.
Eq.~(\ref{Glauber2}) is equivalent to the similar expression for surface scattering in the eikonal approximation developed in Ref.~[\cite{Bortolani}].

The thermal average in Eq.~(\ref{Glauber2}) can be readily carried out in the harmonic approximation, and after taking proper account of the commutation relations of the displacement operators at different times and positions the result is~\cite{Maradudin}
\begin{eqnarray} \label{Glauber3}
w({\bf k}_f,{\bf k}_i) ~=~ \frac{1}{\hbar^2} ~ \left(\frac{\hbar^2 k_{fz}}{mL}\right)^2
~ \left|\left(  \Phi_f(z) \left| \hat{T}_{z}  \right|\Phi_i(z)\right) \right|^2
~ \int_{-\infty}^{+\infty} \, d t ~
e^{i (E_i-E_f)t/\hbar} ~
\\ \nonumber
\times ~~
\frac{1}{L^4} ~ \int ~d {\bf R} ~ \int ~d {\bf R^\prime} ~
e^{-i {\bf K} \cdot ({\bf R}-{\bf R^\prime})} ~
e^{-i \delta k_z [\xi({\bf R}) -\xi({\bf R^\prime}) ]}
~ e^{-2W(\Delta{\bf  k})}
~e^{2 \mathcal{W}(\Delta{\bf  k}; {\bf R},{\bf R^\prime},t)}
~.
\end {eqnarray}
The argument of the Debye-Waller factor $\exp\{-2W \}$ is given by
\begin{eqnarray} \label{2W}
2W(\Delta{\bf  k}) ~=~ \left\langle
(\Delta{\bf  k} \cdot {\bf u})^2
  \right\rangle
~,
\end {eqnarray}
and a standard approximation is to assume that all cross terms in Eq.~(\ref{2W}) average to zero and that the mean square displacements in all three mutually perpendicular cartesian directions are the same.  This leads to
\begin{eqnarray} \label{2W1}
2W(\Delta{\bf  k}) ~\longrightarrow~
\Delta{\bf  k}^2
\left\langle
  { u}_z^2
  \right\rangle
~=~
\frac{3 \hbar^2 \Delta{\bf  k}^2 T_S }{M_C k_B \Theta_D^2}
~.
\end {eqnarray}

The position and time dependent correlation function appearing in the exponential of
Eq.~(\ref{Glauber3}) is given by
\begin{eqnarray} \label{C1}
2 \mathcal{W}(\Delta{\bf  k}; {\bf R},{\bf R^\prime},t) ~=~
\left\langle ~
 \Delta{\bf  k} \cdot {\bf u}({\bf R},t) ~
\Delta{\bf  k} \cdot {\bf u}({\bf R^\prime},0)
  ~\right\rangle
~.
\end {eqnarray}
If the surface is periodic the correlation function also exhibits this periodicity and can be expanded in terms of the normal  modes of vibration and appears as~\cite{Maradudin,Manson-91}
\begin{eqnarray} \label{C2}
2 \mathcal{W}(\Delta{\bf  k}; {\bf R},{\bf R^\prime},t) ~=~
\sum_{\alpha,\alpha^\prime=1}^3 ~ \Delta k_\alpha \Delta k_\alpha^\prime ~
\sum_{{\bf Q}, \nu}  ~ \frac{\hbar}{2 N_C M_C \omega_\nu({\bf Q}) } ~
\\ \nonumber
\times~~
e_\alpha({\bf Q}, \nu) ~ e_{\alpha^\prime}({\bf Q}, \nu) ~
e^{i {\bf Q} \cdot ({\bf R} - {\bf R^\prime})} ~
\left\{  \left[ n_\nu({\bf Q})  +1  \right] ~ e^{-i \omega_\nu({\bf Q}) t} ~+~
n_\nu({\bf Q}) ~ e^{i \omega_\nu({\bf Q}) t}
\right\}
~,
\end {eqnarray}
where $ e_\alpha({\bf Q}, \nu)$ is the $\alpha$ cartesian component of the polarization vector of the mode with parallel wave vector $ {\bf Q}$ and perpendicular quantum number $\nu$ ($\nu$ is discrete for surface modes and is a continuous variable for the bulk modes corresponding to the perpendicular wave vector) and $N_C$ is the number of modes.
The mode frequency is $ \omega_\nu({\bf Q})$ and $ n_\nu({\bf Q})$ is the Bose-Einstein function
\begin{eqnarray} \label{BE}
n_\nu ({\bf Q}) ~=~ \frac{1}{e^{\frac{\hbar \omega_\nu ({\bf Q})}{k_B T_S}} -1}
~.
\end {eqnarray}
Comparison of the correlation function of Eq.~(\ref{C1}) or~(\ref{C2}) and the Debye-Waller exponent of Eq.~(\ref{2W}) shows that
\begin{eqnarray} \label{BE1}
W(\Delta{\bf  k}) ~=~
\mathcal{W}(\Delta{\bf  k}; {\bf R} ={\bf R^\prime},t=0)
~,
\end {eqnarray}
a fact that becomes important later when discussing the classical multiphonon limit of the transition rate.

Eq.~(\ref{Glauber3}) is the starting point for developing and discussing the various regimes of scattering that can occur in fast atom diffraction from surfaces.  These regimes range from the purely quantum mechanical case to completely incoherent classical scattering and in the subsections below we discuss several of these possibilities.

\subsection{Classical Scattering} \label{classical}

The classical regime occurs when the scattering is completely incoherent as for example when a large number of phonons are transferred in the collision and the quantum coherence of the incident beam is completely destroyed.  One case in which this occurs is when the Debye-Waller factor becomes negligibly small implying that its argument $2W$ is large.~\cite{Manson-91,Weare}
The value of the Debye-Waller agrument is essentially a measure of the number of phonons transferred in the collision, and it will be large when either $\Delta {\bf k}$ or the mean square displacement (or effectively the temperature $T_S$) becomes large.  When the Debye-Waller argument $2W$ becomes large the correlation function $2 \mathcal{W}$ of Eq.~(\ref{C2}) also becomes large and furthermore its most important contributions come from the region of small $t$ and small ${\bf R} - {\bf R^\prime}$.   In this case the leading terms in the expansion of $2 \mathcal{W}$ are
\begin{eqnarray} \label{C3}
2 \mathcal{W}(\Delta{\bf  k}; {\bf R},{\bf R^\prime},t) ~\approx~
2 W(\Delta {\bf k})
\\ \nonumber
~-~ i t ~  \sum_{\alpha,\alpha^\prime=1}^3 ~ \Delta k_\alpha \Delta k_\alpha^\prime ~
\sum_{{\bf Q}, \nu}  ~ \frac{\hbar }{2 N_C M_C  } ~
e_\alpha({\bf Q}, \nu) ~ e_{\alpha^\prime}({\bf Q}, \nu) ~
\\ \nonumber
~-~
t^2 ~ \sum_{\alpha,\alpha^\prime=1}^3 ~ \Delta k_\alpha \Delta k_\alpha^\prime ~
\sum_{{\bf Q}, \nu}  ~ \frac{\hbar \omega_\nu({\bf Q})}{4 N_C M_C  } ~
e_\alpha({\bf Q}, \nu) ~ e_{\alpha^\prime}({\bf Q}, \nu) ~
  \left[ 2 n_\nu({\bf Q})  +1  \right]
\\ \nonumber
~-~
 \sum_{\alpha,\alpha^\prime=1}^3 ~ \Delta k_\alpha \Delta k_\alpha^\prime ~
\sum_{{\bf Q}, \nu}  ~ \frac{\hbar [{\bf Q} \cdot ( {\bf R} - {\bf R^\prime})]^2}{4 N_C M_C  \omega_\nu({\bf Q})} ~
e_\alpha({\bf Q}, \nu) ~ e_{\alpha^\prime}({\bf Q}, \nu) ~
  \left[ 2 n_\nu({\bf Q})  +1  \right]
~,
\end {eqnarray}
where the Debye-Waller argument $2W(\Delta {\bf k})$ is the same as that of Eq.~(\ref{2W})

The simplest approximation for evaluating the expansion of Eq.~(\ref{C3}) is to assume a Debye-Model with bulk-like symmetry in which all cross terms vanish and for this case it simplifies to
\begin{eqnarray} \label{C4}
2 \mathcal{W}(\Delta{\bf  k}; {\bf R},{\bf R^\prime},t) ~\approx~
2 W(\Delta {\bf k})  -i \Delta E_0 t
-  \Delta E_0  k_b T_S t^2
- \frac{\Delta E_0  k_b T_S ( {\bf R} - {\bf R^\prime})^2 }{2 \hbar^2 v_R^2}
~,
\end {eqnarray}
where the classical recoil energy is given by $\Delta E_0 = \hbar^2 \Delta {\bf k}^2/2 M_C $,   and the velocity parameter $v_R$ is given by the relation~\cite{Brako}
\begin{eqnarray} \label{vR}
\frac{1}{v_R^2} ~=~  \frac{1}{k_B T_S}
 \sum_{\alpha,\alpha^\prime=1}^3 ~ \Delta \hat{k}_\alpha \Delta \hat{k}_\alpha^\prime ~
\sum_{{\bf Q}, \nu}  ~ \frac{\hbar ({\bf Q} \cdot  \hat{{\bf R}})^2}{2 N_C   \omega_\nu({\bf Q})} ~
e_\alpha({\bf Q}, \nu) ~ e_{\alpha^\prime}({\bf Q}, \nu) ~
  \left[ 2 n_\nu({\bf Q})  +1  \right]
~,
\end {eqnarray}
where $ \hat{{\bf R}} $ is a unit vector in the direction of ${\bf R} - {\bf R^\prime} $.
The parameter $v_R$ of Eq.~(\ref{vR}) is a weighted average of phonon speeds parallel to the surface and it can be evaluated for several simple models of the surface phonon density~\cite{Brako}.  These models produce values that are of order of the bulk acoustic phonon velocities or the Rayleigh wave velocity.

With the simple evaluation of the correlation function of Eq.~(\ref{C4}) the transition rate of Eq.~(\ref{Glauber2}) can be evaluated, a process equivalent to a steepest descents approximation.  Ignoring for the moment the final term involving the spatial dependence in Eq.~(\ref{C4}) the result is
\begin{eqnarray} \label{W3}
w({\bf k}_f,{\bf k}_i) ~=~ \frac{2 \pi}{\hbar} \left(\frac{\hbar^2 k_{fz}}{mL}\right)^2
~ \left|\left(  \Phi_f(z) \left| \hat{T}_{z}  \right|\Phi_i(z)\right) \right|^2
~ \left|  A({\bf K}) \right|^2 ~
\\ \nonumber \times~~~
\sqrt{\frac{1}{4 \pi k_B T_S \Delta E_0}} ~
\exp\left\{ - \frac{\left(E_f - E_i  + \Delta E_0  \right)^2}{4 k_B T_S \Delta E_0} \right\}
~.
\end {eqnarray}

Eq.~(\ref{W3}) is an expression that describes a single classical collision.  For example, if the prefactor transition matrix elements are chosen to be a constant, i.e., if
\begin{eqnarray} \label{P1}
 \left(\frac{\hbar^2 k_{fz}}{mL}\right)^2
~ \left|\left(  \Phi_f(z) \left| \hat{T}_{z}  \right|\Phi_i(z)\right) \right|^2
~ \left|  A({\bf K}) \right|^2 ~
=~ C
~,
\end {eqnarray}
the choice appropriate for hard sphere scattering where the classical cross section is uniform in all angular directions, then it becomes the classical transition rate well-known in neutron scattering~\cite{Sjolander}.  It is also an expression that has been demonstrated to explain single collision events in low energy alkali ion scattering from metal surfaces~\cite{Muis-96,Powers-04} as well as energy-resolved measurements of rare gas scattering from molten metal surfaces.~\cite{Muis-97}

The spatial dependence appearing in Eq.~(\ref{C4}) that was ignored in deriving the transition rate of Eq.~(\ref{W3}) will become of importance to the discussion below in Sec.~\ref{spa}.  However, it is of interest to note here that
another classical expression for the transition rate was obtained by Brako and Newns under the assumption of a smooth surface with thermal vibrations.  If all terms of Eq.~(\ref{C4}), including the term in the spatial dependence, are used to evaluate Eq.~(\ref{Glauber2}), but at the same time assuming that the surface is flat, i.e., setting the corrugation function $\xi({\bf R}) $ equal to a constant, the result is~\cite{Brako}
\begin{eqnarray} \label{W4}
w({\bf k}_f,{\bf k}_i) ~ \propto ~
\sqrt{\frac{1}{4 \pi k_B T_S \Delta E_0}} ~ \frac{1}{ 4 \pi k_B T_S \Delta E_0}
\\ \nonumber \times ~~
\exp\left\{ - \frac{\left(E_f - E_i  + \Delta E_0  \right)^2 + 2 \hbar^2 v_R^2 {\bf K}^2}
{4 k_B T_S \Delta E_0} \right\}
~,
\end {eqnarray}
where  the transition matrix prefactors have again been set equal to a constant.
This expression has also been shown to describe rare gas scattering from metal surfaces.~\cite{Muis-99,Hayes-07}
For the conditions considered here for fast atom scattering at grazing incidence conditions
Eq.~(\ref{W4}) will not be a useful approximation.  For the expected values of $v_R$, which is  in the neighborhood of phonon speeds, the Gaussian-like function in parallel momentum will be very broad and will not have an appreciable effect on the overall transition rate.

Thus, the basis for describing classical scattering will be
Eq.~(\ref{W3}), but as mentioned above it describes a single scattering event in which the projectile particle is assumed to make a transition from its initial state of momentum
$\hbar {\bf k}_i$ to the final state of momentum $\hbar {\bf k}_f$.  However, the discussion in Appendix~\ref{multi} shows that this is not at all the case.  Instead a grazing angle projectile will follow a trajectory along the surface and will encounter successive, incoherent collisions with approximately $N$ atoms before being deflected into the final state of momentum $\hbar {\bf k}_f$.

Under grazing angle conditions, which imply  small total scattering angles $\theta$, the transition rate of Eq.~(\ref{W3}) becomes a very sharp Gaussian as a function of final energy $E_f$ with an average energy loss given by
\begin{eqnarray} \label{El}
E_i~ - ~E_f ~\approx ~ \Delta E_0 ~\approx~ \mu \, E_i \, \theta^2
~,
\end {eqnarray}
where $\mu$ is the mass ratio and the final term on the right of Eq.~(\ref{El}) arises because in a small angle collision with little energy loss $ |\Delta {\bf k}| \approx k_i \theta$.   This same result can be obtained from the Baule conditions for scattering through small angles as shown in Appendix~\ref{multi}.

As shown in more detail in Appendix~\ref{multi} the multiple scattering events along the trajectory can be viewed as a series of $N$ convolutions of the Gaussian-like differential reflection function.  The energy loss during these $N$ collisions is $N$ times that of Eq.~(\ref{El}) but with the total scattering angle $\theta$ replaced by the fractional angle suffered at each of the $N$ collisions which is approximately $\theta/N$.  Given that the square of the full width at half maximum (FWHM) of the Gaussian approximation to Eq.~(\ref{W3}) is proportional to $ 4 k_B T_S \Delta E_0$ and that for the convolution of  Gaussians the squared width of the convolution is the sum of the squares of the individual widths, the final result is a transition rate that looks exactly like Eq.~(\ref{W3}) except that the recoil energy is replaced by
\begin{eqnarray} \label{El2}
\Delta E ~=~ \frac{\Delta E_0}{N} ~=~ \frac{\mu \, E_i \, \theta^2}{N}
~.
\end {eqnarray}
As shown in Appendix~\ref{multi} this energy loss is very small and, moreover, is nearly entirely associated with the motion in the slow directions.  The energy loss in the fast direction is negligible in comparison because it can be readily shown that
$\Delta k_x^2$ is smaller than $\Delta k_z^2 + \Delta k_y^2$ by a factor of $\theta_i^2$ which for grazing angles gives a reduction of several orders of magnitude.
This small energy loss in the fast direction implies that the scattered projectiles will lie close to the energy conservation circle defined by
\begin{eqnarray} \label{circ1}
k_{fz}^2 + k_{fy}^2 ~=~  k_{iz}^2 + k_{iy}^2
\end {eqnarray}
which also implies that the total scattering angle $\theta$ is given in all cases to a good approximation by $2 \theta_i$.

Combining the above simple approximation of the total scattering angle with the evaluation of $N=6/\Gamma a \theta_i$  for the exponentially repulsive potential of Eq.~(\ref{ex}) as obtained in Appendix~\ref{multi} leads to
\begin{eqnarray} \label{El3}
\Delta E  ~=~ \frac{4 \mu \, E_i \, \theta_i^2}{N} ~=~ \frac{2}{3} ~ \mu E_i \Gamma a \theta_i^3
~,
\end {eqnarray}
where $a$ is the distance between surface atoms in the fast direction.

This then leads to a final expression for the classical transition rate which looks similar to Eq.~(\ref{W3}) but with important modifications: (1) $\Delta E_0$ is replaced by the much smaller  $\Delta E$, (2) the scattering amplitude $\left|  A({\bf K}) \right|^2$ separates into the product of $ \left|  A(\Delta k_x) \right|^2 ~ \left|  A(\Delta k_y) \right|^2$.  However, in the fast direction the surface is smoothly averaged and  $ \left|  A(\Delta k_x) \right|^2$ is a constant.  For the slow direction parallel to the surface
$\left|  A(\Delta k_y) \right|^2$ is the probability of being deflected out of the scattering plane.  In the classical limit it will contain the rainbow features of the surface profile, which for example, can be calculated from the classical limit of the eikonal approximation of Eq.~(\ref{A1}).
(3) Finally, the transition matrix element in the normal direction is given by the lognormal distribution of
Eq.~(\ref{A22}) with $\overline{\theta_f}$ given by the relation for the energy conservation  circle of Eq.~(\ref{circ1}).
\begin{eqnarray} \label{W4a}
w({\bf k}_f,{\bf k}_i)= \frac{2 \pi}{\hbar} \left(\frac{\hbar^2 k_{fz}}{mL}\right)^2
P(\theta_f)
\left|  A(\Delta k_y) \right|^2
\sqrt{\frac{1}{\pi k_B T_S \Delta E}}
\exp\left\{ - \frac{\left(E_f - E_i  + \Delta E  \right)^2}{4 k_B T_S \Delta E} \right\}
.
\end {eqnarray}
The relationship between the transition rate and the differential reflection coefficient actually measured in the experiments is made explicit in Appendix~\ref{DRC}.

The differential reflection coefficient described by Eq.~(\ref{W4a}) has a width in the final polar angle (or equivalently in the direction of $k_{fz}$) dictated by the width of the lognormal distribution $P(\theta_f)$.  Its spread in the azimuthal direction (or equivalently the $k_{fy}$ direction) is determined by the form factor
$\left|  A(\Delta k_y) \right|^2$ which depends on the shape of the surface potential along the $y$-direction.  The intensity will exhibit peaks at the rainbow angles which correspond to the inflection points in the surface profile and the spread in the azimuthal direction will be limited to the classically allowed region, which is typically within the limits of the rainbow features allowed by  the largest tilt angle of the surface profile.

\subsection{Purely Quantum Mechanical Scattering} \label{quantum}

The case of purely quantum mechanical scattering in all three dimensions is formally no different from the case of ordinary atom-surface diffraction at low energies.  This is most easily discussed in terms of the eikonal approximation~\cite{Bortolani} or the kinematical approximation~\cite{Manson-91} in which case the transition rate becomes the same as Eq.~(\ref{Glauber3}) except with
$ \left|\left(  \Phi_f(z) \left| \hat{T}_{z}  \right|\Phi_i(z)\right) \right|^2$ taken to be constant.
\begin{eqnarray} \label{Glauber4}
w({\bf k}_f,{\bf k}_i) ~=~ \frac{1}{\hbar^2} ~ \left(\frac{\hbar^2 k_{fz}}{mL}\right)^2
~ \int_{-\infty}^{+\infty} \, d t ~
e^{i (E_i-E_f)t/\hbar} ~
\\ \nonumber
\times ~~
\frac{1}{L^4} ~ \int ~d {\bf R} ~ \int ~d {\bf R^\prime} ~
e^{-i {\bf K} \cdot ({\bf R}-{\bf R^\prime})} ~
e^{-i \delta k_z [\xi({\bf R}) -\xi({\bf R^\prime}) ]}
~ e^{-2W(\Delta{\bf  k})}
~e^{2 \mathcal{W}(\Delta{\bf  k}; {\bf R},{\bf R^\prime},t)}
~.
\end {eqnarray}
This is then usually developed into a series ordered in numbers of phonons transferred by expanding the exponential of the correlation function.  The elastic scattering is the zeroth order term in this series given by
\begin{eqnarray} \label{Glauber5}
w({\bf k}_f,{\bf k}_i) ~=~ \frac{1}{\hbar^2} ~ \left(\frac{\hbar^2 k_{fz}}{mL}\right)^2
~ \int_{-\infty}^{+\infty} \, d t ~
e^{i (E_i-E_f)t/\hbar} ~
\\ \nonumber
\times ~~
\left| \frac{1}{L^2} ~ \int ~d {\bf R} ~
e^{-i {\bf K} \cdot {\bf R}} ~
e^{-i \delta k_z \xi({\bf R})} \right|^2
~ e^{-2W(\Delta{\bf  k})}
~.
\end {eqnarray}
The integral over time becomes a simple $\delta(E_f-E_i)$ stating the conservation of energy.  If the corrugation function $\xi({\bf R})$ is periodic in both directions parallel to the surface the eikonal approximation matrix element reduces to an integral over a single unit cell multiplied by a Kronecker $\delta$-function restricting the parallel momentum transfer to reciprocal lattice vectors ${\bf G}$ of the surface
\begin{eqnarray} \label{Glauber6}
w({\bf k}_f,{\bf k}_i) ~=~ \frac{2 \pi}{\hbar} ~ \left(\frac{\hbar^2 k_{fz}}{mL}\right)^2
\sum_{{\bf G}} ~ |A({\bf G})|^2
~ e^{-2W(\Delta{\bf  k})}
~\delta_{{\bf K},{\bf G}} ~ \delta(E_f-E_i)
~,
\end {eqnarray}
where the scattering amplitude is
\begin{eqnarray} \label{Glauber7}
A({\bf G}) ~=~  \frac{1}{S_{uc}} ~ \int_{uc} ~d {\bf R} ~
e^{-i {\bf G} \cdot {\bf R}} ~
e^{-i \Delta k_z \xi({\bf R})}
~.
\end {eqnarray}

As stated above, Eq.~(\ref{Glauber6}) is the eikonal approximation result for diffraction from a two-dimensional periodic and strongly repulsive surface.  However, in general if $\hbar^2 k_{fz} A({\bf G})/mL$ is replaced by the transition matrix element of the full elastic interaction potential Eq.~(\ref{Glauber6}) becomes exact and is not an approximation.

There are several reasons why it is unlikely that the purely quantum mechanical diffraction described by the case of Eq.~(\ref{Glauber6}), other than possibly a peak at the specular position,  will be observed in the $k_{fx}$ direction in fast atom diffraction:
(1) for energies of hundreds of eV the Debye-Waller factor would be be extremely small except for the most grazing incident conditions, while experimentally it is known that diffraction features in the transverse $k_{fy}$-direction are  observed for energies into the keV range at incident angles of the order of one degree above the surface plane.~\cite{Roncin-07,Winter-07}
(2) The experimental evidence indicates that in the fast direction the atomic projectile experiences an averaged potential that is not corrugated, implying that the matrix elements will be small for non-zero order diffraction peaks in the $ k_{fx}$-direction.~\cite{Salin}
(3) The conditions of energy and parallel momentum conservation determine the value of the perpendicular momentum and similar to Eq.~(\ref{circ}) above it is restricted to the Laue circle given by
\begin{eqnarray} \label{circ2}
k_{fz}^2 ~=~ k_{iz}^2 ~-~ {\bf G}^2 ~-~ 2 {\bf G} \cdot {\bf K}_i
~.
\end {eqnarray}
With $k_{ix} >>k_{iz}$ the cross term on the right hand side of of Eq.~(\ref{circ2}) will dominate and for positive $G_x$ values $ k_{fz}^2 <0$ and those beams will be evanescent.~\cite{Roncin-07}  For negative $G_x$ values  $k_{fz} $ will become substantially larger than  $k_{iz}$ implying that the final diffraction angle will be larger than $\theta_i$ and will not satisfy grazing conditions for the exiting beam.  This again leads to the conclusion that such diffraction features with $G_x \neq 0$ would have negligible intensity.

\subsection{Quantum Behavior in only Two Dimensions} \label{1D}

A more likely scenario for producing diffraction is that the problem reduces to a mixed quantum-classical case, with purely quantum mechanical behavior in the two slow directions (the $z$- and $y$-directions) and classical behavior in the fast direction.   In this case the motion of the atomic projectile in the fast direction is regarded as a series of $N$ incoherent and successive collisions with the surface atoms just as in Section~\ref{classical} above meaning that the lognormal distribution in final angles is retained.  However, the quantum treatment of Section~\ref{quantum} above is carried out for the $y$-direction.  The result is
\begin{eqnarray} \label{Glauber8}
w({\bf k}_f,{\bf k}_i) ~=~ \frac{2 \pi}{\hbar} ~ \left(\frac{\hbar^2 k_{fz}}{mL}\right)^2
~ P(\theta_f) ~
\sum_{ G_y} ~ |A( G_y)|^2
~ e^{-2W(\Delta{\bf  k})/N}
~\delta_{\Delta k_y, G_y} ~ \rho(E_f)
~,
\end {eqnarray}
where $ \rho(E_f)$ is the density of final states in translational energy identified in Sec.~\ref{spa} below.
The  scattering amplitude in the $y$-direction calculated with the eikonal approximation is now
\begin{eqnarray} \label{Glauber9}
A(G_y) ~=~  \frac{1}{a_y} ~ \int_{0}^{a_y} ~d y ~
e^{-i G_y y} ~
e^{-i \Delta k_z \xi(y)}
~,
\end {eqnarray}
which for a simple one-dimensional sinusoidal corrugation function
$  \xi(y) ~=~ h a_y \sin\left(\frac{2 \pi}{a_y} y \right)$ where $h$ is the corrugation strength
leads to the well-known Bessel function solution\cite{Levi-77}
\begin{eqnarray} \label{E1Da}
 A( G_y) ~=~  J_{|m|}(h a_y \Delta k_{z})
~,
\end {eqnarray}
where $J_n(q)$ is the ordinary Bessel function of order $n$ and argument $q$.

An important modification has been applied to the Debye-Waller factor, it appears divided by a factor of $N$.  If the interaction with the surface is considered to be a series of incoherent collisions with $N$ surface sites, a Debye-Waller factor should be associated with each of these collisions.  Such a multiplicative cumulation of Debye-Waller factors is used in LEED for describing the multiple collisions that the electrons incur as they penetrate the surface.~\cite{Duke}
However, the Debye-Waller argument $2W$ for each of these small collisions is smaller than that of  Eq.~(\ref{2W1}) by a factor of $1/N^2$ for reasons similar to those that were discussed in connection with Eq.~(\ref{El2}) above, i.e., because
$ |\Delta {\bf k}| \approx k_i \theta$ and for each small collision
$\theta \approx 2 \theta_i/N$.
Thus, this argument implies that the Debye-Waller factor of Eq.~(\ref{Glauber8}) is transformed according to
\begin{eqnarray} \label{DWx}
e^{-2W} ~\longrightarrow~ \left( e^{-2W/N^2} \right)^N ~=~ e^{-2W/N}
~.
\end {eqnarray}

This same result can be arrived at by a somewhat different argument.~\cite{Rousseau}  The Debye-Waller factor is the average of the phase induced by the thermal displacement
\begin{eqnarray} \label{DWx1}
e^{-W} ~=~ \left\langle  e^{i \Delta {\bf k}\cdot {\bf u} } \right\rangle
~=~  e^{-\frac{1}{2} \left\langle ( \Delta {\bf k}\cdot {\bf u} )^2\right\rangle}
~.
\end {eqnarray}
If the total phase is considered to be the sum of the individual phases picked up in each of the small collisions
\begin{eqnarray} \label{DWx2}
\Delta {\bf k}\cdot {\bf u} ~ \approx ~ \sum_{n=1}^N ~ \Delta {\bf k}_n\cdot {\bf u}_n
~,
\end {eqnarray}
and if the individual scattering events are considered uncorrelated the sum of all cross terms in the mean square thermal average $ \left\langle ( \Delta {\bf k}\cdot {\bf u} )^2\right\rangle$ will go to zero leading to a sum containing $N$ mean square phases.  Assuming a similar phase for each small collision again results in the same expression as in Eq.~(\ref{DWx}).

There is a seemingly apparent conflict inherent in the result of Eq.~(\ref{Glauber8}) because the density of states in final energy $\rho(E_f) $, if evaluated strictly, is a $\delta$-function in the difference between final and initial translational energy.
However, if the motion in the fast direction is classical then the translational energy of the projectile cannot be conserved because some energy will be exchanged with the surface through the classical collisions.  The conservation of energy is very nearly preserved because the exchange of energy associated with the fast direction is negligibly small for the same reasons as explained above in Section~\ref{classical}, i.e., as long as all angles are grazing $\Delta k_x^2$ is smaller than the momentum transfer in the other two directions by a factor of order $\theta_i^2$.  Thus the condition of energy conservation is weakly relaxed, and this allows for the range of
$ k_{fz}$ values implied by final polar angle width of the lognormal distribution.  The expression for $\rho(E_f)$ that allows for a small exchange of energy is determined in
 Eq.~(\ref{k1}) of Section~\ref{spa} below.

The scattered intensity spectrum described by  Eq.~(\ref{Glauber8}) exhibits a number of interesting features.  In the $k_{fy}$ direction there are narrow diffraction peaks whose intensity are given by the transition matrix element of the potential which in the eikonal approximation is  Eq.~(\ref{Glauber9}).  In the actual experiment, the widths in this direction would be limited by the transverse coherence length  of the experiment.  In the $k_{fz}$ direction these diffraction features are broad with a width determined by the angular spread of the lognormal distribution.  All of these diffraction features lie roughly along the Laue circle given by
\begin{eqnarray} \label{circ3}
k_{fz}^2 ~=~ k_{iz}^2 ~-~ { G}_y^2 ~-~ 2  G_y k_{iy}
~,
\end {eqnarray}
or equivalently
\begin{eqnarray} \label{k2}
\cos\theta_f ~=~ \cos \theta_i ~ \frac{\cos \phi_i}{\cos \phi_f}
~.
\end {eqnarray}
In each of the diffraction features $k_{fz}$ increases from the bottom to the top of the broad streak which means that the intensity can vary considerably as is seen from the phase variation in the scattering amplitude of  Eq.~(\ref{Glauber9}).  This quantum interference effect can be either destructive or constructive, and when it is destructive it can even make portions of the streak disappear.  This quantum interference effect can also cause a given streak to appear as if it were shifted above or below the Laue circle.  Finally, the Debye-Waller factor is considerably larger than that implied by the simple expression of  Eq.~(\ref{2W1}) making it possible to observe diffraction peaks with
measurable intensities.
The Debye-Waller factor essentially depends on momentum transfer only in the quantum mechanical directions, i.e., $\Delta {\bf k}^2 \approx \Delta k_y^2 + \Delta k_z^2$ because at such small scattering angles $\Delta k_x^2$ is negligible in comparison.
All of these features are observed in the experimental measurements~\cite{Roncin-07,Winter-07}.

\subsection{Near-Classical Scattering with Spatial Correlations}  \label{spa}

The two cases above in Secs.~\ref{quantum} and \ref{1D} discuss  conditions in which quantum mechanical diffraction peaks can arise in the scattered intensity.   There is yet a third case in which spatial correlations will give rise to diffraction features in an otherwise nearly classical scattered spectrum.  This case is similar to the case of diffraction features observed in the multiphonon background in electron diffraction experiments such as RHEED where they are known as Kikuchi lines.~\cite{Kikuchi}

The starting point for this treatment is Eq.~(\ref{Glauber3}) assuming an averaged flat potential in the fast direction and a corrugation function $\xi(y)$ of periodic length $a_y$ in the $y$-direction
\begin{eqnarray} \label{Glauber10}
w({\bf k}_f,{\bf k}_i) ~=~ \frac{1}{\hbar^2} ~ \left(\frac{\hbar^2 k_{fz}}{mL}\right)^2
~ P(\theta_f) ~
~ \int_{-\infty}^{+\infty} \, d t ~
e^{i (E_i-E_f)t/\hbar} ~
\\ \nonumber
\times ~~
\frac{1}{L^2} ~ \int_{-\infty}^{+\infty} ~d {y} ~ \int_{-\infty}^{+\infty} ~d {y^\prime} ~
e^{-i {\Delta k_y (y-y^\prime)}} ~
e^{-i \delta k_z [\xi({y}) -\xi({y^\prime}) ]}
~ e^{-2W(\Delta{\bf  k})}
~e^{2 \mathcal{W}(\Delta{\bf  k}; {y},{y^\prime},t)}
~.
\end {eqnarray}
The periodicity of the corrugation function is expressed as
 $ \xi(y + n a_y) = \xi(y)$ with $n$  an integer
and the fact that the correlation function $\mathcal{W}$ depends only on the difference in displacement $y-y^\prime$ allows one to replace the spatial integrals by a sum of integrals over a single unit cell.  Since the correlation function has very weak spatial dependence over the distance of a single period (or equivalently, under the assumption that each unit cell vibrates rigidly) this leads to
\begin{eqnarray} \label{Glauber10a}
w({\bf k}_f,{\bf k}_i) ~=~ \frac{1}{\hbar^2} ~ \left(\frac{\hbar^2 k_{fz}}{mL}\right)^2
~ P(\theta_f) ~
~ \int_{-\infty}^{+\infty} \, d t ~
e^{i (E_i-E_f)t/\hbar} ~
\\ \nonumber
\times ~~
\frac{1}{L^2} ~ \int_{0}^{a_y} ~d {y} ~ \int_{0}^{a_y} ~d {y^\prime} ~
e^{-i {\Delta k_y (y-y^\prime)} }~
e^{-i \delta k_z [\xi({y}) -\xi({y^\prime}) ]}
~ e^{-2W(\Delta{\bf  k})}
\\ \nonumber
\times ~~
\sum_{n=-\infty}^{+\infty} ~ \sum_{n^\prime=-\infty}^{+\infty}~
e^{-i {\Delta k_y a_y (n-n^\prime) }}
~e^{2 \mathcal{W}(\Delta{\bf  k}; {na_y}-{n^\prime a_y},t)}
~.
\end {eqnarray}
Making the classical limit expansion of the correlation function  as in Eqs.~(\ref{C3}) and~(\ref{C4}) leads to
\begin{eqnarray} \label{Glauber10b}
w({\bf k}_f,{\bf k}_i) ~=~ \frac{1}{\hbar^2} ~ \left(\frac{\hbar^2 k_{fz}}{mL}\right)^2
~ P(\theta_f) ~
~ \int_{-\infty}^{+\infty} \, d t ~
e^{i (E_i-E_f)t/\hbar} ~
~ e^{-i \Delta E t
-  \Delta E  k_b T_S t^2}
\\ \nonumber
\times ~~
\left| A(\Delta k_y)   \right|^2
\sum_{n=-\infty}^{+\infty} ~ \sum_{n^\prime=-\infty}^{+\infty}~
e^{-i {\Delta k_y a_y (n-n^\prime) }} ~
e^{- {\Delta E  k_b T_S ( {n} - {n^\prime})^2 a_y^2} / {2 \hbar^2 v_R^2}}
~.
\end {eqnarray}
where $\Delta E_0$ has been replaced by $\Delta E$ of Eq.~(\ref{El3}) to account for the multiple collisions with $N$ surface unit cells along the fast direction, and
as in Eq.~(\ref{Glauber9})
\begin{eqnarray} \label{Glauber9a}
A(\Delta k_y) ~=~  \frac{1}{a_y} ~ \int_{0}^{a_y} ~d y ~
e^{-i \Delta k_y y} ~
e^{-i \Delta k_z \xi(y)}
~.
\end {eqnarray}
The discrete sums reduce to a single summation given by
\begin{eqnarray} \label{Glauber11}
\frac{L}{a_y}~ \sum_{n=-\infty}^{+\infty} ~
e^{-i {\Delta k_y a_y n }} ~
e^{- {\Delta E  k_b T_S n^2 a_y^2} / ({2 \hbar^2 v_R^2} ) }
~\approx ~
\\ \nonumber
\frac{L}{a_y}~ \frac{\hbar v_R}{\sqrt{2 \pi \Delta E k_B T_S}} ~
\sum_{G_y} ~ e^{2 \hbar^2 v_R^2 (\Delta k_y-G_y)^2 / (4 k_B T_S \Delta E)}
~,
\end {eqnarray}
where $L$ is a quantization length parallel to the surface.
Clearly, this summation is a periodic function in
$\Delta k_y \longrightarrow \Delta  k_y+G_y$
where $G_y = 2 \pi \nu / a_y$ with $\nu$ an integer, and the right hand side of
 Eq.~(\ref{Glauber11}) becomes a good approximation if
$ \Delta E  k_B T_S  a_y^2 / (2 \hbar^2 v_R^2) < 1 $.

The time integral in Eq.~(\ref{Glauber10}) is the same Gaussian integral encountered in Section~\ref{classical} above so the final result for the transition rate is
\begin{eqnarray} \label{Glauber12}
w({\bf k}_f,{\bf k}_i) ~=~ \frac{2 \pi}{\hbar} ~ \left(\frac{\hbar^2 k_{fz}}{mL}\right)^2
~ P(\theta_f) ~
\frac{1}{\sqrt{4 \pi \Delta E k_B T_S}} ~
\exp\left\{  - \frac{(E_f - E_i + \Delta E)^2}{4 \pi \Delta E k_B T_S}   \right\}
\\ \nonumber
\times ~~
\left| A(\Delta k_y)   \right|^2 ~
\frac{\hbar v_R}{\sqrt{2 \pi \Delta E k_B T_S}} ~
\sum_{G_y} ~ e^{2 \hbar^2 v_R^2 (\Delta k_y-G_y)^2 / (4 k_B T_S \Delta E)}
~.
\end {eqnarray}

Eq.~(\ref{Glauber12}) describes periodic diffraction features lying roughly centered on the Laue circle of Eq.~(\ref{circ3}).  In the $k_{fy}$ direction these features are Gaussian-like with a width parameter $k_0$ given by
\begin{eqnarray} \label{k0}
k_0 ~=~ \sqrt{\frac{ \Delta E  k_b T_S  a_y^2} {2 \hbar^2 v_R^2} }
~.
\end {eqnarray}
In the $k_{fz}$ direction they appear as long streaks whose width is governed by the lognormal distribution in final polar angles. The strength of these peaks at any point
$( k_{fz},k_{fy})$ in the wave vector space perpendicular to the fast direction is given by the  form factor $ \left| A(\Delta k_y)   \right|^2$ which in this treatment is approximated by the eikonal expression of
Eq.~(\ref{Glauber9a}).  This form factor can vary substantially over the $k_{fz}$ extent of the diffraction feature, just as discussed above in connection with Section~\ref{1D} even to the point of making parts of a diffraction feature disappear because of destructive quantum interference arising from the corrugation within the unit cell of the potential.  There is no Debye-Waller factor associated with these diffraction features.  Instead the temperature dependence varies inversely with a power of the temperature, in this case $T_S^{-3/2}$,  typical of classical scattering.  Also typical of classical scattering, the widths of all peaks increases as $\sqrt{T_S}$ for temperatures large enough where zero point motion is not important.  The energy spread within these peaks is quite narrow and given by
\begin{eqnarray} \label{k1}
\rho\left(  E_f  \right) ~=~
\frac{1}{\sqrt{4 \pi \Delta E k_B T_S}} ~
\exp\left\{  - \frac{(E_f - E_i + \Delta E)^2}{4 \pi \Delta E k_B T_S}   \right\}
~.
\end {eqnarray}
which approaches the energy $\delta$-function $\delta\left(  E_f - E_i +\Delta E  \right)$ for small $\Delta E$.
Eq.~(\ref{k1}) also identifies the density of final energy states
$ \rho\left(  E_f  \right)$ introduced above in Eq.~(\ref{Glauber8}).


\section{Calculations}\label{calcs}

In the following a few calculations are presented to illustrate some of the different cases presented in the sections above.  The system considered is $^3$He scattering from LiF(001) along the $\langle 110  \rangle$ direction with an incident energy of 200 eV and an incident polar angle of $2^\circ$ with respect to the surface.  With the LiF lattice vector of $b=4.03$ \AA~the effective reciprocal lattice vector is $G_y =  2 \pi \sqrt{2}/b = 2.21$ \AA$^{-1}$.

For the calculations the surface  Debye temperature is chosen to be 530 K (a value corresponding to a bulk Debye temperature of 750 K divided by $ \sqrt{2}$), the mass $M_C= 19$ amu corresponding to that of a fluorine atom, the exponential potential range parameter is $\Gamma = 2.0$ \AA$^{-1}$, the surface temperature is 300 K and the velocity parameter is chosen to be $v_R=4000$ m/s which is the Rayleigh wave speed of LiF(001).~\cite{Rayleigh}
The  scattering amplitude $A(\Delta k_y)$ that determines the relative intensities of the diffraction features is chosen to be the eikonal Bessel function approximation of Eq.~(\ref{E1Da}) for a simple one-dimensional sinusoidal corrugation function with the arbitrary choice $h=0.023$.
The evaluation of the number $N=6/ \Gamma a \theta_i$ of LiF lattice sites encountered along the path of the He atom during the interaction is approximately $N \sim 21$ for these conditions.

\begin{figure}
\includegraphics[width=6.5in]{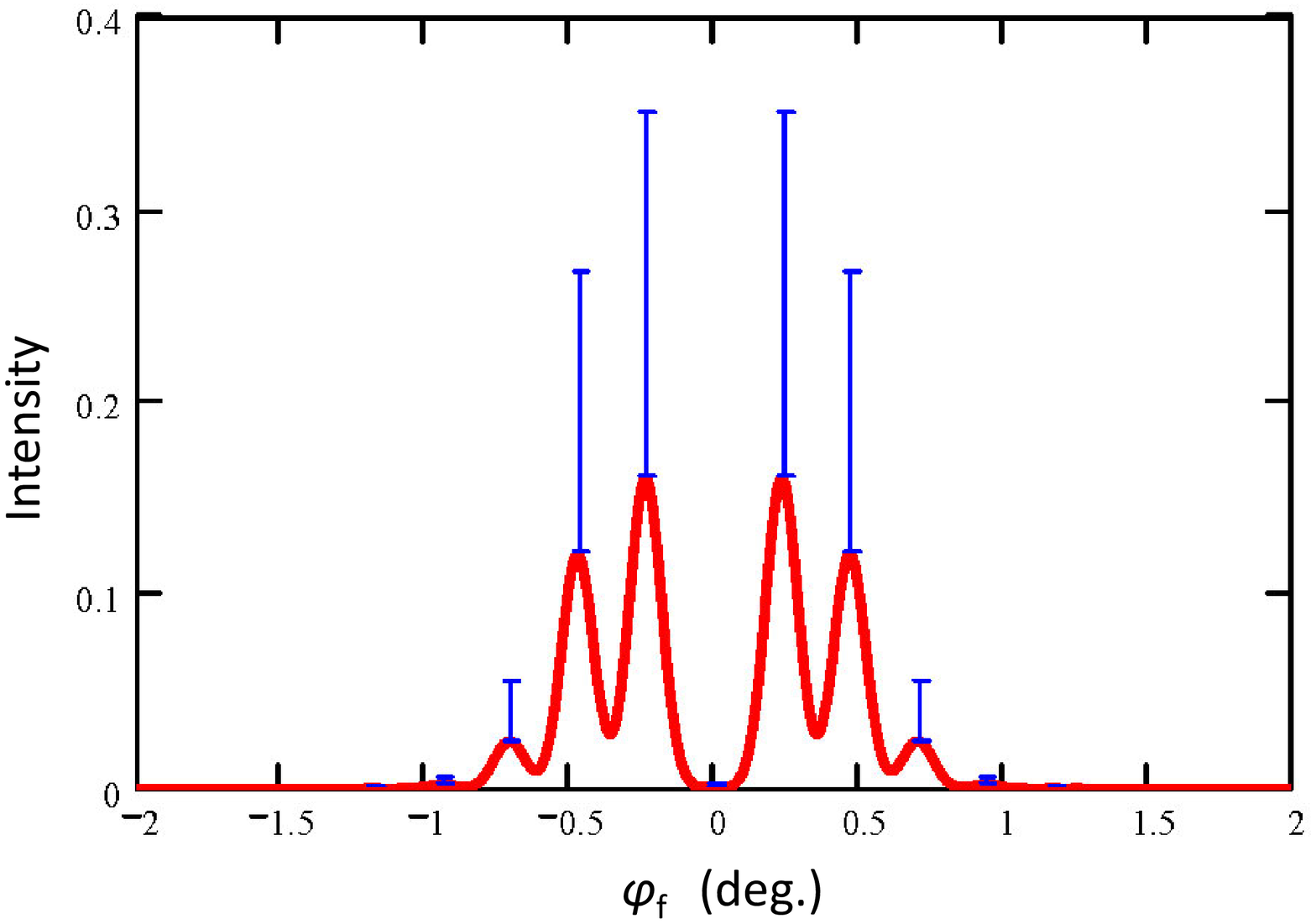}
\caption{The scattered intensity as a function of $\phi_f$ with $\theta_f$ lying on the Laue circle for a 200 eV beam of $^3$He incident on LiF(001)$<110>$ at an incident angle $\theta_i=2^\circ$ with respect to the surface.  All other parameters are specified in the text.  The broad features are the calculations for the nearly classical case of Eq.~(\protect\ref{Glauber12}) and the sharp vertical lines at the diffraction peak positions represent the intensities of the quantum diffraction peaks calculated from Eq.~(\protect\ref{Glauber8}).
}
\label{fig1}
\end{figure}

Fig.~\ref{fig1} shows a graph plotting the differential reflection coefficient $dR/d\theta_f d\phi_f$ as a function of $\phi_f$ (measured in degrees)  with the polar angle chosen to lie along the Laue circle of Eq.~ (\ref{circ3}) or~(\ref{k2}).
The intensity plotted on the vertical axis is measured in arbitrary units.

The broad diffraction features centered about the diffraction points are calculations of the nearly classical transition rate of Eq.~(\ref{Glauber12}). The sharp vertical lines located precisely at the diffraction positions $ \Delta k_y = G_y$ are the quantum diffraction peaks calculated from Eq.~(\ref{Glauber8}).  The relative intensity of the two sets of calculations with respect to each other is arbitrarily chosen.  In this particular example the zeroth order or specular diffraction peak at $ \phi_f = 0$ has nearly zero intensity and peaks of up to order 4 are visible on either side of the specular position.

\begin{figure}
\includegraphics[width=6.5in]{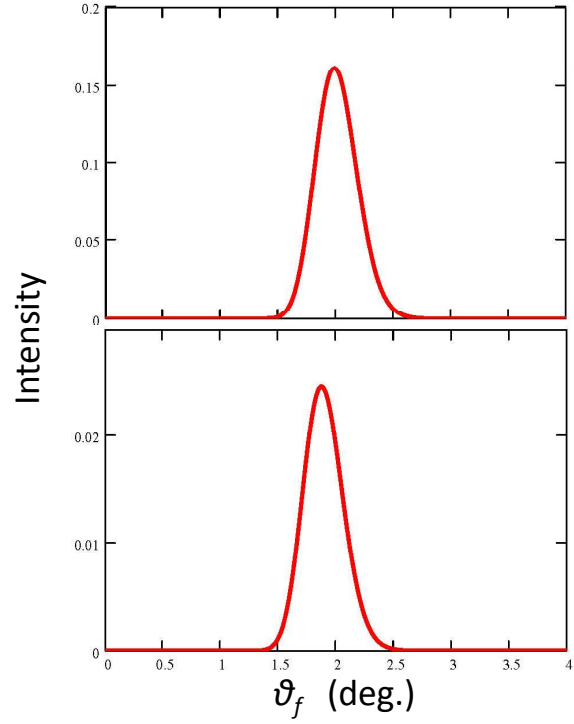}
\caption{For the same conditions as Fig.~\protect\ref{fig1} the differential reflection coefficient is plotted as a function of $\theta_f$ for $\phi_f$ fixed at the first and third order diffraction peak position in the upper and lower panels, respectively.
}
\label{fig2}
\end{figure}

Fig.~\ref{fig2} shows the diffraction feature calculated with the nearly classical transition  rate of Eq.~(\ref{Glauber12}) but this time plotted as a function of $\theta_f$ measured in degrees for  values of $\phi_f$ fixed at the positions of the first and third order diffraction peaks in the upper and lower panels, respectively.  This shows a rather broad feature with a FWHM of about half a degree which is essentially the width of the lognormal distribution in final polar angle.  The polar angle distribution for the diffraction peaks calculated with the quantum model of Eq.~(\ref{Glauber8}) would be identical.  Even for the third order diffraction peak a small shift  of the distribution along the Laue circle to polar angles smaller than $\theta_f=2^\circ$  is evident.

\begin{figure}
\includegraphics[width=6.5in]{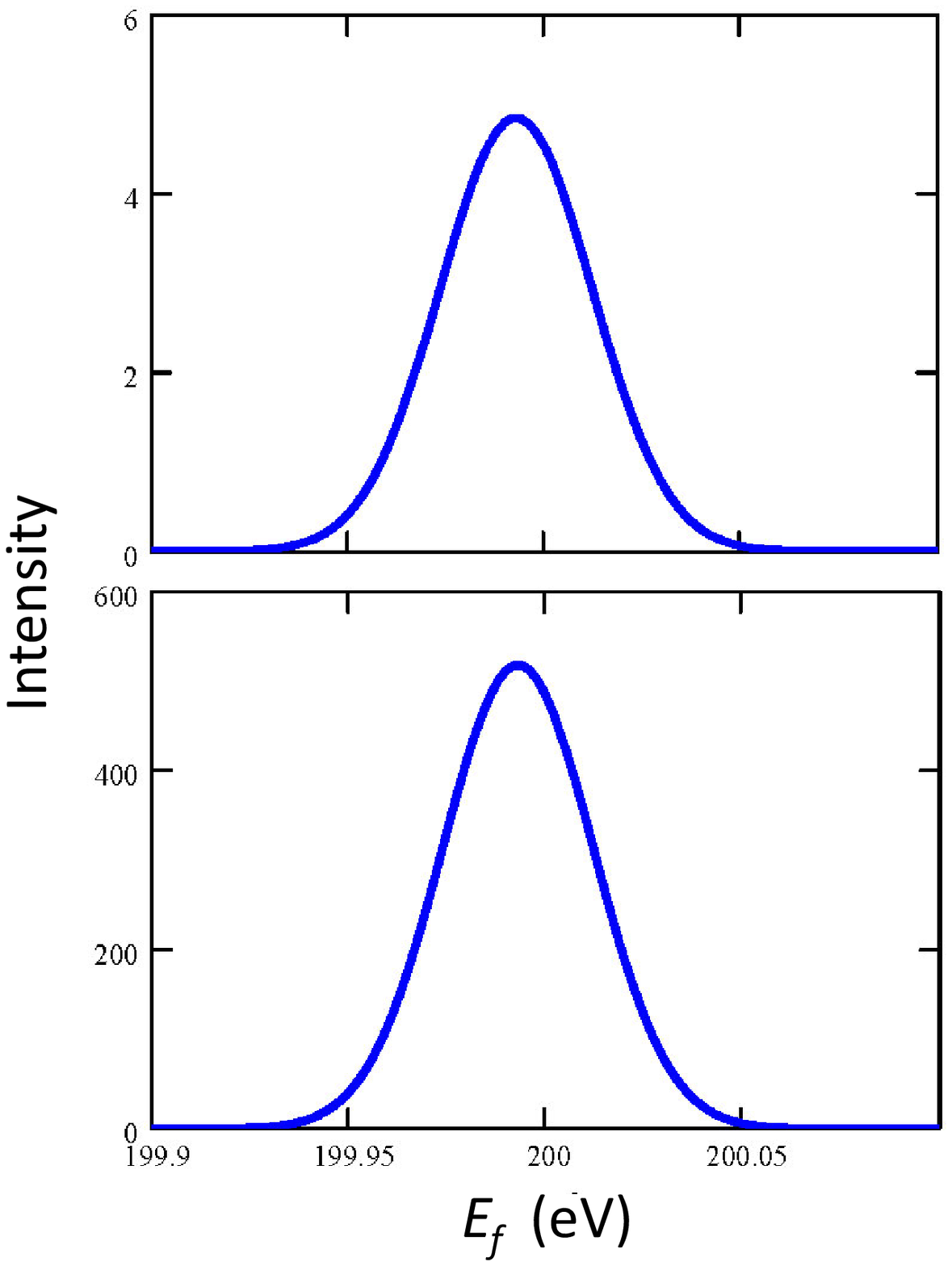}
\caption{The energy-dependent differential reflection coefficient of the nearly classical model of Eq.~(\protect\ref{Glauber12}) is plotted as a function of final energy $E_f$ at the specular and second order diffraction peak positions in the upper and lower panels, respectively.
}
\label{fig4}
\end{figure}

 Figs.~\ref{fig1} and~\ref{fig2} show the intensity integrated over all final energies because this is what is experimentally measured by the channel plate detector.  In order to illustrate the energy dependence the full three-dimensional differential reflection coefficient
$d R ({\bf k}_f,{\bf k}_i) / d E_f d \Omega_f$ is plotted as a function of final energy $E_f$ in Fig.~\ref{fig4}  for final angles at the specular position and at the second order diffraction peak in the upper and lower panels, respectively.  The calculations are carried out with the nearly classical
transition  rate of Eq.~(\ref{Glauber12}).
The two peaks are quite similar, and in fact the energy spectrum in the vicinity of all diffraction peaks looks very much the same.
This figure demonstrates the remarkably small energy losses experienced by these projectiles in spite of the high incident energy and relatively large incident angle.
The full width at half maximum is less than 50 meV and the
most probable final energy is approximately 10 meV smaller than the 200 eV incident energy, or a fractional energy loss of about $5 \times 10^{-5}$.  This energy loss is in agreement with the value of about 7.2 meV given by the expression of Eq.~(\ref{El3}).


\section{Conclusions}\label{conclusion}

The objective of this paper is to discuss the variety of scattering regimes that can occur in the diffraction of fast atomic and molecular projectiles at grazing angles from surfaces.  Such a discussion should begin with mention of purely quantum diffraction in all three spatial dimensions.  This would be essentially no different than a description of the well-known technique of ordinary He atom diffraction at low (thermal) energies, and is unlikely to apply to the present experiments because of the high energies and correspondingly extremely large velocities in the fast direction.

However, here we review the salient features of three regimes of diffraction and scattering that appear to be possible.  These are (A) quantum behavior in two dimensions, (B) near-classical scattering with spatial correlations, and in addition, (C) possible contributions to the incoherent background are briefly reviewed.  Each of these regimes has very characteristic features in their dependence on the experimentally controllable variables.

\subsection{Quantum behavior in two dimensions} \label{Conc1}

In this case, discussed in detail above in Sec.~\ref{1D}, the motion in the fast direction, parallel to the surface, is classical while the motion normal to the surface and also that parallel to the surface but transverse to the fast direction is quantum mechanical.  In spite of the classical motion with multiquantum energy transfer in one of the three dimensions, quantum features can be observed in the other dimensions because the energy losses in the fast direction are exceedingly small, too small to cause a complete loss of quantum coherence.

This regime is characterized by diffraction features that appear as streaks in the $k_{fz}$-direction, with the length of the streaks governed by the lognormal distribution $P(\theta_f)$ in final polar angles $\theta_f$.  In the $k_{fy}$-direction these peaks are narrow, their widths being limited by the instrument resolution.  These peaks are observed to be distributed roughly around the Laue circle whose wave vector components are given by Eq.~(\ref{circ2}).

The intensity dependence of these diffraction features obeys a variant of the classic Debye-Waller behavior.  The temperature dependence of the peak intensity is an exponentially decreasing function of temperature divided by a power of $T_S$, and in the treatment shown in  Eq.~(\ref{Glauber8}) it is seen that power is one.  The argument of the Debye-Waller factor is considerably reduced over the standard value that would be calculated according to Eq.~(\ref{2W1}) using the initial and final momenta of the diffraction features.  The $2W$ argument of the Debye-Waller factor is effectively divided by $N$, the number of collision sites along the path of motion.  It is this effect that allows the Debye-Waller factor to be large enough that the diffraction features are measurable.  Also, the motion in the fast direction does not contribute appreciably to the momentum transfer in the Debye-Waller argument of Eq.~(\ref{2W1}) and this is the cause of the extremely small energy losses in that dimension.  For beams incident along high symmetry directions of the surface, where the Laue circle of Eq.~(\ref{circ2}) appears on the channel plate detector as symmetric about the specular spot, it is of interest to note that the Debye-Waller exponent $2W$ gets smaller, and hence the Debye-Waller factor becomes larger, with increasing $G_y$ reciprocal lattice vectors, and this is because of the associated decrease of the final perpendicular momentum transfer.  If the incident beam is slightly miss-aligned with respect to a high symmetry direction, which would imply a small non-vanishing ${\bf K}_i$ in Eq.~(\ref{circ2}), the Laue circle becomes skewed with respect to the specular point and the preceeding statement becomes true only for those reciprocal lattice vectors $G_y$ whose corresponding perpendicular momentum transfers decrease, since some perpendicular momentum transfers may increase.  The temperature dependence of the width of the diffraction features differs in one respect from the standard Debye-Waller behavior.  In the $k_{fy}$-direction, the widths are independent of temperature, which is the normal D-W behavior.  However, the widths of the streaks in the $k_{fz}$-direction increases in proportion to $\sqrt{T_S}$.  This increase in width comes from the lognormal distribution in Eq.~(\ref{Glauber8}) which defines the overall length of the streaks.  It is also of interest to note that the lognormal distribution shows that the width of the streaks in the $k_{fz}$-direction will be proportional to $\Gamma$, the range parameter of the exponentially decreasing interaction potential in the surface-normal direction.

The relative intensities of the diffraction features associated with different $G_y$ reciprocal lattice vectors, as well as the relative intensity within the streak of the diffraction feature due to a single given $G_y$, is governed by the corrugation landscape of the potential as expressed in the transition matrix amplitudes $A(G_y) $ of Eq.~(\ref{Glauber8}).  For example, the corrugation of the surface can produce rainbow effects and supernumerary rainbow effects in the diffraction peak distributions, as has been recently reported~\cite{Winter-08}.  Intensity modulation due to the shape of the potential can also be observed within the streak of a single diffraction feature and such effects can cause the streak to appear "shifted" above or below its expected position on the Laue circle, as is apparent in the observations~\cite{Roncin-07}.

In addition to the above responses of the diffraction features to the controllable experimental variables, it is of interest to discuss several possibilities that might be open for future experimental investigation.  The first of these is the unlikely possibility of observing quantum diffraction effects in the high velocity direction, due to reciprocal lattice vectors in the $G_x$-direction.  It is of interest to note the reasons why this is unlikely.  For the case of positive $G_x$ (meaning the direction of $G_x$ is parallel to the fast velocity component) at the energies used in the present experiments the corresponding $k_{fz}$ components given by the Laue equation of (\ref{circ2}) become rapidly evanescent and hence unobservable.  On the other hand, for negative $G_x$ the value of $k_{fz}$ becomes sufficiently large that the diffraction features are projected several degrees in polar angle above the specular position.  This means that these peaks would also be unobservable because a flat ordered surface cannot scatter intensity into such large scattering angles.

A possible effect closely related to supernumerary rainbows is the observation of quantum interference due to scattering of terraces at different atomic height intervals on the surface as seen in thermal energy He atom diffraction~\cite{Lapujoulade} and in reflection high energy electron diffraction~\cite{Kikuchi}.  This phenomenon should also manifest itself in the present fast atom diffraction-type experiments, although the effect may be limited by the relatively short coherence length of the experiment.

An effect observed very early in the history of He atom diffraction at thermal energies was the observation of selective adsorption, or resonances with the bound states of the potential~\cite{Stern,Devonshire}.  In principle these resonances, which are characterized by sharp increases or decreases in diffraction peak intensities as an experimentally controllable parameter is varied through resonance conditions, should be present in the current experiments.  The simplest statement of the condition for selective adsorption resonance is that the value of $\hbar^2 k_{fz}^2 / 2 m$ must be negative (corresponding to an evanescent beam) and nearly equal to the energy of a bound state of the potential.  In the present experiments, the simplest way to look for selection adsorption events would be by slightly varying the incident azimuthal angle $\phi_i$ away from a high symmetry direction. When the incident beam is aligned along a high symmetry direction with $\phi_i=0^\circ$ the diffraction features observed on the channel plate detector lie roughly along the Laue circle of Eq.~(\ref{circ2}) which appears symmetric with respect to the specular spot.  However, if $\phi_i$ deviates slightly from zero, for example upon a small azimuthal rotation of the target crystal, the observed Laue circle becomes skewed and asymmetric.  Along with this asymmetric shifting of the Laue circle, many of the low-index diffraction features will become evanescent and have $k_{fz}$ values that pass through conditions for resonance.

One of the most important uses of thermal energy He atom diffraction has been the measurement of single phonon transfers and surface phonon dispersion relations.  Because of the large disparities of energy between that of typical phonons and the kinetic energies used in theses experiments it is unlikely that most single surface phonon inelastic events will be measurable. However, it is worth mentioning that there is one type of inelastic process that could be measured by these experiments, and this would be energy exchange with dispersionless excitations, e.g., surface phonons, surface plasmons, surface polaritons, etc. whose energies do not depend on momentum transfer parallel to the surface.  For example, diffraction should still be observable even in the event that the atomic projectiles exchange one or more dispersionless inelastic quanta because such processes would not destroy the quantum coherence of the system.

\subsection{Near Classical Scattering with Spatial Correlations}  \label{Concl2}

This case is described in detail in Sec.~\ref{spa} and is analogous to Kikuchi lines observed in RHEED due to simultaneous multiple phonon transfer and diffraction.  The diffraction features appear similar to those due to two-dimensional diffraction described in the above section, except that there is no Debye-Waller factor and there is temperature dependent broadening in both the $k_{fz}$ and $k_{fy}$ directions.  This type of diffraction could appear under conditions where the 2-D quantum behavior of the above section has negligible intensity, or the two can appear simultaneously as shown in the calculations presented in Fig.~\ref{fig1}.  The expression for the transition rate is given in Eq.~(\ref{Glauber12})  and the major difference with the 2-D diffraction case of Eq.~(\ref{Glauber8}) is that the product of the Debye-Waller factor and the discrete delta-function dependence in parallel momentum transfer have been replaced by a periodic density of states consisting of broader peaks.  The physical reason for these differences is the fact that multiple phonon quanta are transferred in this case, and the spatial region of coherence from which an interference diffraction pattern is built is no longer infinitely large, but becomes finite in extent.

The diffraction features are still manifest as relatively long streaks in the $k_{fz}$-direction whose length is governed by the lognormal distribution and these streaks appear aligned roughly along the Laue circle.  In the $k_{fy}$-direction the peaks are now broadened beyond the instrumental limit, and the temperature dependence of the FWHM of this broadening is the same in both directions, i.e., proportional to $\sqrt{T_S}$.  The temperature dependence of the maximum intensity of the diffraction features, as shown in Eq.~(\ref{Glauber12}), is inversely proportional to $T^{3/2}$.

\subsection{Classical Incoherent Limit}  \label{concl3}

The classical incoherent limit is essentially a part of the background intensity that should be observed in the experiments.  In this paper two different mechanisms from which it can arise are discussed.  The first of these, presented in Sec.~\ref{classical}, arises from scattering from regions of the surface whose heights are randomly distributed but whose average interaction potential is a repulsive exponential with range parameter $\Gamma$.  The appearance of this background intensity is a feature broadened in the $k_{fz}$-direction by the lognormal distribution, and broad in the $k_{fy}$-direction according to the extent permitted by the associated classical scattering form factor $|A(\Delta k_y)|^2$.  The temperature dependence of the maximum of this background intensity varies approximately inversely with the surface temperature, while the FWHM is
proportional to $\sqrt{T_S}$. Because the peak width in the polar angle $\theta_f$ is governed mainly by the lognormal function, this width is only weakly dependent on incident energy.  The peak intensity also is inversely proportional to $\Gamma \sqrt{\mu}$, while the FWHM is directly proportional to this same quantity.

The total energy loss is evaluated in Appendix~\ref{expo} in the classical limit of large incident energy and high temperatures and is given by $\Delta E= 2 \mu E_i \Gamma a \theta_i^3/3$.  This energy loss, for the present grazing-angle geometry with small $\theta_i$ is negligible and roughly five orders of magnitude smaller than the incident energy.  Most of this energy loss can be ascribed to motion in the direction perpendicular to the surface, and it can be readily shown that the the fraction of energy that can be ascribed to diminishing the fast $\hbar k_{fx}$ component of momentum is smaller than the above $\Delta E$ by a factor proportional to $\theta_i^2$.  It is of interest to note that $\Delta E$ is proportional to the range parameter of the repulsive exponential potential, implying that measurement of the energy losses over a range of incident energies, or when combined with measurements of intensity and FWHM dependence, may provide a means to measure this important physical parameter.

The other, and radically different, type of background expected to be measured in high energy atomic scattering experiments is that due to binary collisions with topological defects such as adatoms or step edges on the surface.
This contribution is described by Eqs.~(\ref{W3}) taken together with the constant form factor of (\ref{P1}). This is discussed further starting with Eq.~(\ref{M2}) of Appendix~\ref{multi}.  This type of binary scattering is spread over a broad range of energies and angles, with a FWHM proportional to $\sqrt{E_i T_S}$ and an intensity at its peak maximum inversely proportional to the same quantity, and its general features under the assumption of hard-sphere scattering are outlined in Eqs.~(\ref{M3}) to~(\ref{M6}).  In addition to its larger energy and angular widths, this type of background can be distinguished from that discussed in the two paragraphs above by its large energy losses, the fractional energy loss under the grazing angle conditions of the present experiments would be approximately $\mu E_i \theta_i^2$ as opposed to the cubic behavior in $\theta_i$ discussed above for $\Delta E$.  This type of classical incoherent scattering is well-documented and has been extensively measured in experiments in low-energy ion scattering over energy ranges similar to those of interest here~\cite{Powers-04}.


\newpage
\appendix
\section{Classical motion in an exponential potential}  \label{expo}

For a point particle moving in a one-dimensional exponentially repulsive potential
\begin{eqnarray} \label{A1a}
V(z) ~=~ V_0 ~ e^{-\Gamma z}
~,
\end {eqnarray}
the expression of conservation of energy is
\begin{eqnarray} \label{A2}
E_z ~=~ \frac{1}{2}m v_z^2 ~+~ V_0 ~ e^{-\Gamma z}
~,
\end {eqnarray}
where $E_z$ is the energy associated with the normal motion and the normal velocity is
\begin{eqnarray} \label{A3}
v_z(t) ~=~ \frac{dz}{dt} ~=~ \sqrt{\frac{2 E_z}{m}   \left(1-e^{-\Gamma (z-z_0)}  \right)}
~,
\end {eqnarray}
where $z_0 = \ln(E_z/V_0)/ \Gamma $ is the classical turning point.
Equation~(\ref{A3}) admits an analytic solution that can be parametrically expressed in terms of $v_{iz}= \sqrt{2 E_z / m} $ as
\begin{eqnarray} \label{A4}
z(t) ~=~ z_{0} ~+~ v_{iz} t ~+~ \frac{2}{\Gamma} ~ \ln \left( \frac{1 ~+~ e^{-\Gamma v_{iz} t}}{2}          \right)
~,
\end {eqnarray}
 as can be readily verified by direct differentiation.

\subsection{Energy loss in a classical grazing incidence collision}

Beginning from the known expression for the energy transfer in a two-body collision with a small total scattering angle $\theta$
\begin{eqnarray} \label{A5}
\Delta E_0 ~=~ E_i - E_f ~=~ \mu \theta^2 E_i
~,
\end {eqnarray}
where $\mu$ is the mass ratio, the small fractional energy loss occurring over a short distance parallel to the surface during which the particle is deflected by the small angle $\delta \theta$ is
\begin{eqnarray} \label{A6}
\delta E ~=~  \mu ( \delta \theta)^2 E_i ~ = ~
\mu E_i \left(  \frac{\delta \theta}{\delta t}  \right)^2  (\delta t)^2
~.
\end {eqnarray}
The assumption of grazing angle scattering also implies that  the  parallel velocity $v_x >> v_z$ implying that
\begin{eqnarray} \label{A7}
\frac{\delta \theta}{\delta t} ~\approx ~  \frac{\delta v_z(t)}{\delta t} ~\frac{1}{v_x}
~ \approx ~ \frac{\delta^2 z(t)}{(\delta t)^2} ~\frac{1}{v_x}
~,
\end {eqnarray}
and for a surface of small periodic length $a$ one can approximate
\begin{eqnarray} \label{A8}
v_x ~=~ \frac{dx}{dt}  ~\approx~ \frac{a}{\delta t}
~.
\end {eqnarray}
Taking the infinitesimal limit, Eq.~(\ref{A6}) becomes
\begin{eqnarray} \label{A9}
\delta E ~=~  \mu ~E_i ~a ~ \left(  \frac{d^2 z(t)}{dt^2} \right)^2 ~ \frac{1}{v_x^3} ~ dt
~.
\end {eqnarray}
Equation~(\ref{A9}) identifies the power loss spectrum, which can be evaluated from Eq.~(\ref{A4}) as~\cite{Roncin,Villette}
\begin{eqnarray} \label{A10}
\frac{dE(t)}{dt} ~=~ \mu ~E_i~ a ~ \frac{1}{v_x^3} ~ \frac{\Gamma^2 v_0^4}{4 \cosh^4(\Gamma v_0 t/2)}
~.
\end {eqnarray}
Integrating Eq.~(\ref{A10}) over all times along the path of motion and recognizing that the total scattering angle for the grazing-angle collision is $\theta ~ \approx ~ 2 \theta_i$ leads to a result that is cubic in the incident angle, rather than the quadratic form suggested by Eq.~(\ref{A5})~\cite{Roncin,Villette}:
\begin{eqnarray} \label{A11}
\Delta E ~=~ \frac{2}{3} ~ \mu E_i \Gamma a \theta_i^3
~.
\end {eqnarray}

The result of Eq.~(\ref{A11}) can be understood in physical terms through the following simple argument: in a grazing-angle collision the interaction can be assumed to be spread roughly equally over $N$ sequential and incoherent collisions with $N$ lattice sites in a line along the surface, and clearly for such grazing-angle conditions $N~ \propto ~ 1/\theta_i$.  If each of these individual collisions has a fractional energy loss equivalent to Eq.~(\ref{A6}) and if the fractional angular deflection is the uniform value
$\delta \theta ~\approx (\theta_i + \theta_f)/N ~ \approx ~ 2 \theta_i/N$ then the sum of all energy losses over the $N$ small collisions along the trajectory is
\begin{eqnarray} \label{A12}
\Delta E ~=~  ~  \frac{4 \mu E_i \theta_i^2}{N}
~.
\end {eqnarray}
Since $N$ is known to be inversely proportional to $\theta_i$, Eq.~(\ref{A12}) gives the same cubic dependence on incident angle as the more exact result of Eq.~(\ref{A11}).  This argument also identifies an approximation for estimating the number of surface cells lying along the path of the collision:~\cite{Roncin,Villette}
\begin{eqnarray} \label{A13}
N ~ \approx ~ \frac{6}{\Gamma a \theta_i}
~.
\end {eqnarray}

\subsection{The lognormal angular distribution}  \label{lognorm}

If the surface potential is exponentially repulsive as in Eq.~(\ref{A1}) and additionally if each element of the surface is subject to a random displacement in the vertical direction the distribution of final scattering angles is lognormal.~\cite{Roncin,Villette}
This is perhaps most easily understood through noting that in a grazing-angle scattering event with $v_{iz} << v_x$ the final angle is approximately given by
\begin{eqnarray} \label{A14}
\theta_f ~ \approx ~ \theta_i ~ \approx ~  \frac{v_{iz}}{v_x}
~ \approx ~  \frac{\sqrt{2 E_z/m}}{v_x}
~,
\end {eqnarray}
and this can be related to the classical turning point through
\begin{eqnarray} \label{A15}
\theta_f ~ \approx ~  \frac{\sqrt{2 V_0/m}}{v_x} ~ e^{- \Gamma z_0 / 2}
~=~ \overline{\theta_f} ~ e^{- \Gamma z_0 / 2}
~.
\end {eqnarray}
Now suppose that the surface is subject to random displacements $u_z$ in the vertical direction whose probability distribution is Gaussian and given by
\begin{eqnarray} \label{A16}
P(u_z) ~=~ \frac{1}{\sqrt{2 \pi \sigma^2}} ~
\exp\left\{ - \frac{u_z^2}{2 \sigma^2}  \right\}
~.
\end {eqnarray}
Even though ultimately the displacements $u_z$ may be due to thermal motion of the surface atoms the motion of the projectile is so fast that they appear static, i.e., the collision time is much shorter than a typical phonon period.

One can then obtain the mean value of $\theta_f$ through
\begin{eqnarray} \label{A17}
\frac{\left\langle \theta_f \right\rangle}{ \overline{\theta_f}} ~=~
\int_{-\infty}^{+\infty} ~ du_z ~ e^{- \Gamma (z_0 - u_z) / 2} ~
\frac{1}{\sqrt{2 \pi \sigma^2}} ~
\exp\left\{ - \frac{u_z^2}{2 \sigma^2}  \right\}
~,
\end {eqnarray}
which gives the result
\begin{eqnarray} \label{A18}
\frac{\left\langle \theta_f \right\rangle}{\overline{\theta_f}} ~=~
e^{\Gamma^2 ~ \sigma^2 / 8}
~.
\end {eqnarray}
If the integration variable  in Eq.~(\ref{A17}) is changed to $\theta_f$ according to
\begin{eqnarray} \label{A19}
\ln \left( \frac{ ~~\theta_f ~~}{\overline{\theta_f}} \right) ~=~
 \frac{\Gamma}{2} (z_0 - u_z)
~,
\end {eqnarray}
the result is
\begin{eqnarray} \label{A20}
 \frac{\left\langle \theta_f \right\rangle}{\overline{\theta_f}}  ~=~
\frac{1}{\Gamma \overline{\theta_f}} \sqrt{ \frac{2}{\pi \sigma^2 }} ~
\int ~ d \theta_f ~
\exp \left\{ - \frac{2 \left[ \ln \left( \frac{ ~~\theta_f~~ }{ \overline{\theta_f} } \right) \right]^2}{ \Gamma^2
  \sigma^2 }    \right\}
~.
\end {eqnarray}
However, relating Eq.~(\ref{A20}) to the standard form in terms of the distribution function $P(\theta_f)$ given in the small angle approximation by
\begin{eqnarray} \label{A21}
 \theta_f  ~=~ \int ~ d \theta_f ~ \theta_f ~ P(\theta_f)
~,
\end {eqnarray}
identifies $P(\theta_f)$ as the lognormal distribution
\begin{eqnarray} \label{A22}
P(\theta_f) ~=~
 \sqrt{\frac{2}{\pi \sigma^2}} ~
\frac{1}{ \Gamma \theta_f} ~
\exp \left\{ -\frac{2 \left[ \ln \left( \frac{~~\theta_f~~}{\overline{\theta_f}} \right) \right]^2}{\Gamma^2
  \sigma^2}    \right\}
~.
\end {eqnarray}

For an ordered surface, the roughness will be caused by thermal motion.  So even though the surface appears static to the rapidly moving projectile, the width of the Gaussian distribution can be associated with the mean square thermal displacement
\begin{eqnarray} \label{A23}
\sigma^2 ~=~ \left\langle u_z^2 \right\rangle
~.
\end {eqnarray}

\section{Convolution of multiple collisions with the surface}  \label{multi}

The differential reflection coefficient describing classical scattering is Eq.~(\ref{W3})
\begin{eqnarray} \label{W3a}
w({\bf k}_f,{\bf k}_i) ~=~  \left(\frac{\hbar^2 k_{fz}}{mL}\right)^2
~ \left|\left(  \Phi_f(z) \left| \hat{T}_{z}  \right|\Phi_i(z)\right) \right|^2
~ \left|  A({\bf K}) \right|^2 ~
\\ \nonumber \times~~~
\sqrt{\frac{\hbar}{4 \pi k_B T_S \Delta E_0}} ~
\exp\left\{ - \frac{\left(E_f - E_i  + \Delta E_0  \right)^2}{4 k_B T_S \Delta E_0} \right\}
~.
\end {eqnarray}
under many conditions, including the case of grazing angle scattering at high energies considered here, this becomes very nearly Gaussian in form leading to a differential reflection coefficient Eq.~(\ref{DRC2}) of the form
\begin{eqnarray} \label{M2}
\frac{d^3R({\bf k}_f,{\bf k}_i)}{d  {E}_f d^2 \Omega_f} ~\approx~
\sqrt{\frac{1}{4 \pi k_B T_S \Delta E_0}} ~
\exp\left\{ - \frac{\left(E_f - E_i  + \Delta E_0  \right)^2}{4 k_B T_S \Delta E_0} \right\}
~.
\end {eqnarray}
When this happens the most probable energy is given by the Baule relation for an elastic two-body collision $\overline{E_f} = f(\mu, \theta) E_i$
where $\mu$ is the mass ratio, $\theta$ is the scattering angle and
\begin{eqnarray} \label{M3}
f(\mu, \theta) ~=~ \left( \frac{\sqrt{1-\mu^2 \sin^2\theta} + \mu \cos \theta}{1+ \mu}  \right)^2
~.
\end {eqnarray}
The recoil energy evaluated at the most probable final energy is $\Delta E_0 = g_{TA}(\mu, \theta) E_i$ with
\begin{eqnarray} \label{M3a}
g_{TA}(\mu, \theta) ~=~ \mu\left( 1 + f(\mu, \theta) - 2 \sqrt{f(\mu, \theta)} \cos \theta \right)
~.
\end {eqnarray}
The mean square energy deviation is $\langle E_f^2 \rangle  = 2 g(\mu, \theta) E_i k_B T_S$
where
\begin{eqnarray} \label{M4}
g(\mu, \theta) ~=~ \frac{g_{TA}(\mu, \theta)}{\left( 1 + \mu - \frac{\mu \cos \theta}{\sqrt{f(\mu, \theta)}}  \right)^2}
~.
\end {eqnarray}
The trajectory approximation to $g(\mu, \theta)$ is $g_{TA}(\mu, \theta) $ and clearly
\begin{eqnarray} \label{M5}
f(\mu, \theta) ~=~ 1 ~-~ g_{TA}(\mu, \theta)
~,
\end {eqnarray}
and for small scattering angles
\begin{eqnarray} \label{M6}
g(\mu, \theta) ~ \approx ~ g_{TA}(\mu, \theta)  ~ \approx ~ \mu \theta^2
~,
\end {eqnarray}

The interaction with the surface is described in terms of  successive collisions with $N$
surface atoms and each of these collisions has a scattering angle which if averaged over the whole path is approximately $\theta_n = 2 \theta_i/N$ since the final scattering intensity is predominantly along the circle of energy conservation.
The differential reflection coefficient for the $n$-th collision is given by
\begin{eqnarray} \label{M7}
\frac{d^3R({\bf k}_n,{\bf k}_{n-1})}{d  {E}_n d^2 \Omega_n} ~\approx~
{\frac{1}{ \sqrt{\pi} \sigma_n}} ~
\exp \left\{ - \frac{\left(E_n -  f_n(\mu, \theta_n) E_{n-1}  \right)^2}{\sigma_n^2} \right\}
~.
\end {eqnarray}
where the definition of $\sigma_n$ is obvious from Eq.~(\ref{M2}).
The differential reflection coefficient resulting from this series of collisions is the convolution of $N$ different successive collisions
\begin{eqnarray} \label{M8}
\frac{d^3R({\bf k}_f,{\bf k}_i)}{d  {E}_f d^2 \Omega_f}  =
\int~ d E_N d \Omega_N ~ \frac{d^3R({\bf k}_f,{\bf k}_N)}{d  {E}_N d^2 \Omega_N}
\cdots
\int ~ d E_1 d \Omega_1   \frac{d^3R({\bf k}_2,{\bf k}_1)}{d  {E}_2 d^2 \Omega_2} `
\frac{dR^3({\bf k}_1,{\bf k}_i)}{d  {E}_1 d^2 \Omega_1}
~.
\end {eqnarray}
The usual rule that the convolution of Gaussians is a Gaussian whose squared width is the sum of the squared widths of each element of the convolution becomes slightly more complicated in this case because of the factor $f_n E_{n-1}$ in the argument.  In this case the final result is
\begin{eqnarray} \label{M8a}
\frac{d^3R({\bf k}_f,{\bf k}_i)}{d  {E}_f d^2 \Omega_f} ~ = ~
{\frac{1}{ \sqrt{\pi} \sigma_T}} ~
\exp\left\{ - \frac{\left(E_n -  f_T E_i  \right)^2}{\sigma_T^2} \right\}
~,
\end {eqnarray}
where
\begin{eqnarray} \label{M9}
f_T ~=~ \prod_{n=1}^N ~ f_n   \longrightarrow ~
1 ~-~  N \mu \left( \frac{2 \theta_i}{N}  \right)^2
~,
\end {eqnarray}
and
\begin{eqnarray} \label{M10}
\sigma_T ~=~ \sum_{n=1}^N \prod_{j=1}^{N-j} ~ f_j^2 ~\sigma_n^2
~ \longrightarrow ~
N \sigma_n^2  ~=~
N \mu \left( \frac{2 \theta_i}{N}  \right)^2 E_i k_B T_S
~,
\end {eqnarray}
where the expressions on the RHS of Eqs.~(\ref{M9}) and~(\ref{M10}) are obtained after making the small angle approximation that all $f_n$ are nearly equal to unity and $\sigma_n$ are small, and both are independent of $n$.
Recalling the definition of $\Delta E$ from  Eq.~(\ref{El3}) the differential reflection coefficient now becomes that of Eq.~(\ref{W4a}):
\begin{eqnarray} \label{M11}
w({\bf k}_f,{\bf k}_i)= \left(\frac{\hbar^2 k_{fz}}{mL}\right)^2
P(\theta_f)
\left|  A(\Delta k_y) \right|^2
\sqrt{\frac{\hbar}{\pi k_B T_S \Delta E}}
\exp\left\{ - \frac{\left(E_f - E_i  + \Delta E  \right)^2}{4 k_B T_S \Delta E} \right\}.
\end {eqnarray}

\section{Experimentally measured quantities}  \label{DRC}

The physical quantity measured by these experiments is the differential reflection coefficient per unit final energy and solid angle denoted as ${dR({\bf k}_f,{\bf k}_i)} / {dE_f d\Omega _f} $.
This is obtained from the transition rate by first forming the differential reflection coefficient per unit volume of wave vector space
\begin{eqnarray} \label{DRC1}
\frac{d^3R({\bf k}_f,{\bf k}_i)}{d  {\bf k}_f} ~=~
\frac{w({\bf k}_f,{\bf k}_i)}{ \Delta k^3 ~ j_i}
~,
\end {eqnarray}
where in terms of the quantization length $L$ in each Cartesian direction the element of volume in phase space is $ \Delta k^3 = (2 \pi/L)^3 $ and the incident flux normal to the surface is $j_i = \hbar k_{iz}/ m L $.
Changing to spherical coordinates according to
$ d  {\bf k}_f = k_f^2 d k_f d \Omega_f$ and recalling that
$E_f = \hbar^2 k_f^2/2 m $ leads to
\begin{eqnarray} \label{DRC2}
\frac{dR^3({\bf k}_f,{\bf k}_i)}{d  {E}_f d^2 \Omega_f} ~=~
\frac{L^4 m^2}{(2 \pi \hbar)^3} ~ \frac{k_f}{k_{iz}} ~
w({\bf k}_f,{\bf k}_i)
~.
\end {eqnarray}

The channel plate detector used in the experiments measures the integral over all final particle energies and thus the measured quantity is the differential reflection coefficient per unit final polar and azimuthal angles which is expressed as
\begin{eqnarray} \label{DRC3}
\frac{d^2R}{d  {\theta}_f d \phi_f} ~=~
\cos(\theta_f) ~ \int_0^\infty ~  d E_f ~
\frac{d^3R({\bf k}_f,{\bf k}_i)}{d  {E}_f d^2 \Omega_f}
~,
\end {eqnarray}
where the polar angle used here is measured with respect to the surface plane, and not the usual one measured with respect to the surface normal.




\newpage
\listoffigures

\newpage


\newpage

\begin{thebibliography}{199}


\bibitem{Roncin-07}  P. Rousseau, H. Khemliche, A. G. Borisov and P. Roncin, Phys. Rev. Lett. {\bf 98}, 016104 (2007).

\bibitem{Winter-07}  A. Sch\"{u}ller, S. Wethekam and H. Winter, Phys. Rev. Lett. {\bf 98}, 016103 (2007).

\bibitem{Winter-08} A. Sch\"uller and H. Winter, Phys. Rev. Lett. {\bf 100}, 097602 (2008).

\bibitem{Burgdorfer-07}  N. Simonovi\'c, F. Aigner, B. Solleder and J. Burgd\"orfer, presented at the ICPEAC conference, Freiburg, Germany (2007).


\bibitem{Weinstock-44}  Robert Weinstock, Phys. Rev. {\bf 65}, 1 (1944).

\bibitem{Glauber-52} R. Glauber, Phys. Rev. {\bf 87}, 189 (1952).

\bibitem{Hove-54} L. Van Hove, Phys. Rev. {\bf 95}, 249 (1954).

\bibitem{Levi-77}  U. Garibaldi, A. C. Levi, R. Spadacini and G. Tommei, Surface Science {\bf 48}, 649 (1975).


\bibitem{Roncin}  Kristel Cordier, Hocine Khemliche and P. Roncin, LCAM-CNRS report, unpublished.

\bibitem{Villette}  J\'er\^ome Villette, Ph. D. Thesis, Universit\'e de Paris-Sud, Orsay, France (2000), unpublished, available online at: http://tel.archives-ouvertes.fr/tel-00106816/en/.

\bibitem{Bortolani}  V. Bortolani and A. C. Levi, Riv. del Nuovo Cimento {\bf 9}, 1 (1986).

\bibitem{MansonSpringer}  J. R. Manson, in {\em Helium Atom Scattering}, E. Hulpke ed., Springer Series in Surface Sciences {\bf 27}, 173 (1992).

\bibitem{Maradudin}  A. A. Maradudin, E. W. Montroll and G. H. Weiss, in {\em Solid State Physics: Theory of Lattice Dynamics in the Harmonic Approximation} (Academic Press, New York, 1963), Suppl. 3).

\bibitem{Manson-91}  J. R. Manson, Phys. Rev. B{\bf 43}, 6924 (1991).

\bibitem{Weare}  John H. Weare, J. Chem. Phys. {\bf 61}, 2900 (1974).


\bibitem{Brako}  R. Brako and D. M. Newns, Phys. Rev. Lett. {\bf 48}, 1859 (1982); R. Brako and D. M. Newns, Surf. Sci. {\bf 117}, 422 (1982).

\bibitem{Sjolander}  A. Sj\"olander, Ark. Phys. {\bf 14}, 315 (1959).

\bibitem{Muis-96} A. Muis and J. R. Manson, Phys. Rev. B {\bf 54}, 2205 (1996).

\bibitem{Powers-04}  J. Powers, J. R. Manson, C. E. Sosolik, J. R. Hampton, A. C. Lavery, and B. H. Cooper, Phys. Rev. B {\bf 70}, 115413 (2004).

\bibitem{Muis-97} 	Andre Muis and J. R. Manson, J.  Chem. Phys. {\bf 107}, 1655 (1997).

\bibitem{Muis-99} A. Muis and J. R. Manson,
J. Chem. Phys. {\bf 111}, 730 (1999).

\bibitem{Hayes-07}  W. W. Hayes and J. R. Manson,  J. Chem. Phys. {\bf 127}, 164714 (2007).


\bibitem{Salin} D. Far\'ias, C. D\'iaz, P. Rivi\`ere, H. F. Busnengo, P. Nieto, M. F. Somers, G. J. Kroes, A. Salin, and F. Mart\'in Phys. Rev. Lett. {\bf 93}, 246104 (2004).


\bibitem{Duke}  C. B. Duke,  and  C. W. Tucker, Jr,  Surf. Sci. {\bf 15}, 231 (1969).

\bibitem{Rousseau}   Patrick Rousseau, Ph. D. Thesis, Universit\'e de Paris-Sud, Orsay, France (2006), unpublished, available online at: http://tel.archives-ouvertes.fr/tel-00106727/en/.


\bibitem{Kikuchi}  A. Ichimiya and P. I. Cohen, {\em Reflection High-Energy Electron Diffraction}, (Cambridge University Press, Cambridge, 2004).


\bibitem{Rayleigh}  G. W. C. Kaye and T. H. Laby, {\em Tables of Physical and Chemical Constants and Some Mathematical Functions}, (Longman, London, 1986).




\bibitem{Lapujoulade} J. Lapujoulade J and Y. Lejay, Journal de Physique Lettres {\bf 38}, L303 (1977).
\bibitem{Stern} I. Estermann and O. Stern, Z. Phys. {\bf 61}, 95 (1930); R. Frisch and O. Stern, Naturwissenschaften {\bf 20}, 721 (1932).

\bibitem{Devonshire}  J. E. Lennard-Jones and A. F. Devonshire, Nature {\bf 137}, 1069 (1936); A. F. Devonshire, Proc. Roy. Soc. London {\bf A 156}, 37 (1936).









\end{thebibliography}
\end{document}